\documentclass[12pt]{article}

\usepackage[dvips]{graphicx}
\usepackage{wrapfig}
\usepackage{amssymb}
\usepackage{amsmath}

\def\IZ{\mathbb {Z}}

\def\IC{\mathbb {C}}
\def\IP{\mathbb {P}}

\renewcommand{\thefootnote}{\fnsymbol{footnote}}

\makeatletter
 \renewcommand{\theequation}{%
       \thesection.\arabic{equation}}
 \@addtoreset{equation}{section}
\makeatother

\makeatletter
\def\eqnarray{%
 \stepcounter{equation}%
 \let\@currentlabel=\theequation
 \global\@eqnswtrue
 \global\@eqcnt\z@
 \tabskip\@centering
 \let\\=\@eqncr
 $$\halign to \displaywidth\bgroup\@eqnsel\hskip\@centering
 $\displaystyle\tabskip\z@{##}$&\global\@eqcnt\@ne
 \hfil$\displaystyle{{}##{}}$\hfil
 &\global\@eqcnt\tw@$\displaystyle\tabskip\z@{##}$\hfil
 \tabskip\@centering&\llap{##}\tabskip\z@\cr}
\makeatother

\begin{document}
\begin{titlepage}

\begin{flushright}
HRI/ST/1301
\end{flushright}

\begin{center}
\vspace*{1cm}
{\Large \bf
Exact K\"ahler Potential for Calabi-Yau Fourfolds}
\vskip 1.5cm
{\large Yoshinori Honma\footnote[1]{yhonma@hri.res.in} and Masahide Manabe\footnote[2]{masahidemanabe@gmail.com}}
\vskip 1.0em
{\it 
Harish-Chandra Research Institute \\
Chhatnag Road, Jhusi, Allahabad 211019, India\\
}
\end{center}
\vskip3cm

\begin{abstract}
We study quantum K\"ahler moduli space of Calabi-Yau fourfolds. Our analysis is based on the recent work by Jockers et al. which gives a novel method to compute the K\"ahler potential on the quantum K\"ahler moduli space of Calabi-Yau manifold. In contrast to Calabi-Yau threefold, the quantum nature of higher dimensional Calabi-Yau manifold is yet to be fully elucidated. In this paper we focus on the Calabi-Yau fourfold. In particular, we conjecture the explicit form of the quantum-corrected K\"ahler potential. We also compute the genus zero Gromov-Witten invariants and test our conjecture by comparing the results with predictions from mirror symmetry. Local toric Calabi-Yau varieties are also discussed.

\end{abstract}
\end{titlepage}


\renewcommand{\thefootnote}{\arabic{footnote}} \setcounter{footnote}{0}

\section{Introduction}

Topological string theory gives us a framework to study ``topological quantum quantities'' as Gromov-Witten invariants corresponding to the worldsheet instanton numbers. Strongly motivated from the string compactification, Calabi-Yau threefolds has been well-studied. The quantum-corrected K\"ahler potential on the K\"ahler moduli space of Calabi-Yau threefold can be determined by a single holomorphic function called prepotential which encodes the information about the genus zero Gromov-Witten invariants. This simplification is due to the special K\"ahler structure of the moduli space of Calabi-Yau threefold \cite{Dixon:1989fj}. Mirror symmetry is an efficient tool to compute the prepotential as first demonstrated in \cite{Candelas:1990rm} to the quintic Calabi-Yau threefold. Although moduli spaces of Calabi-Yau manifolds whose complex dimension is greater than three do not have such a special K\"ahler structure, the higher dimensional mirror symmetry is still useful to compute Gromov-Witten invariants \cite{Greene:1993vm,Mayr:1996sh,Klemm:1996ts} (see also \cite{Klemm:2007in}). As pioneered in \cite{Witten:1993yc}, various Calabi-Yau manifolds can be realized at the IR fixed points of the two dimensional ${\cal N}=(2,2)$ gauged linear sigma model (GLSM). Mirror symmetry for Calabi-Yau manifolds described by abelian GLSMs is well understood \cite{Hosono:1993qy,Hosono:1994ax} and also proved physically in \cite{Hori:2000kt}. However, for general cases with non-abelian gauge groups, comprehensive study of mirror symmetry is yet to come.

Recently, it was proposed in \cite{Jockers:2012dk} that Gromov-Witten invariants of Calabi-Yau manifolds can be computed without using mirror symmetry. More precisely, they proposed the relation
$$
e^{-K}=Z_{\mbox{\scriptsize GLSM}}
$$
between the exact K\"ahler potential $K$ on the quantum K\"ahler moduli space of a Calabi-Yau manifold and the exact partition function $Z_{\mbox{\scriptsize GLSM}}$ of an ${\cal N}=(2,2)$ GLSM on $S^2$ which was computed in \cite{Benini:2012ui,Doroud:2012xw}. This means that we can extract the genus zero Gromov-Witten invariants only from the GLSM calculation. They also checked the consistency of their proposal for some known examples and two physical proofs are given in \cite{Gomis:2012wy}. Their method is applicable to non-abelian GLSMs and the authors of \cite{Jockers:2012dk} also made predictions for Gromov-Witten invariants of a determinantal Calabi-Yau threefold. Note that, as mentioned in \cite{Jockers:2012dk,Gomis:2012wy}, their method has a close relationship to the toric mirror symmetry.

The aim of this paper is to study the quantum K\"ahler moduli space of Calabi-Yau fourfold using the novel method of \cite{Jockers:2012dk}. We conjecture a formula of the exact K\"ahler potential in (\ref{Kahl4}) and provide a prescription to compute the Gromov-Witten invariants of Calabi-Yau fourfolds. We test our conjecture by computing the Gromov-Witten invariants of several Calabi-Yau fourfolds and comparing them with known results in the literature. We also study local toric Calabi-Yau varieties, and propose a correspondence in (\ref{toricclaim}) which precisely express a local toric analogue of the statement of \cite{Jockers:2012dk}.

This paper is organized as follows. In Section 2, we first review the mirror symmetry in general dimension. It gives an introduction to the Gromov-Witten theory from the aspects of mirror symmetry. Then we summarize the proposal of \cite{Jockers:2012dk} with the results of \cite{Benini:2012ui,Doroud:2012xw}. In Section 3, we represent our main result about the quantum K\"ahler moduli space for Calabi-Yau fourfolds, and refer to the relationship to the mirror manifold. In Section 4, we test our proposal by demonstrating the exact GLSM calculation for several examples of compact Calabi-Yau fourfolds. In Section 5, we consider local toric Calabi-Yau varieties and some remarks about the non-compact limit are given. Section 6 is devoted to the conclusion and discussion. We summarize the results of general complete intersections in the Grassmannian in Appendix A.

\section{Some known results}

In this section, we collect some known results on mirror symmetry of higher dimensional Calabi-Yau manifolds \cite{Greene:1993vm,Mayr:1996sh,Klemm:1996ts}. Then we summarize the proposal in \cite{Jockers:2012dk} which gives a new method to compute the exact K\"ahler potential on the quantum K\"ahler moduli space of Calabi-Yau manifolds via two dimensional ${\cal N}=(2,2)$ GLSM partition function.

\subsection{Moduli spaces and mirror symmetry}

As is well known, one can define two types of two dimensional topological field theories, the A-model and the B-model, by gauging either the vector $U(1)_V$ or the axial vector $U(1)_A$ R-symmetries in the ${\cal N}=(2,2)$ non-linear sigma model \cite{Witten:1988xj, Witten:1991zz} (see also \cite{Hori:2000kt}). The A-model on a Calabi-Yau $d$-fold $X$ without boundary depends only on the K\"ahler moduli of $X$ and is independent of the complex structure moduli. On the other hand, the B-model on a Calabi-Yau $d$-fold $X^*$ without boundary only captures the information about the complex structure moduli of $X^*$. By Batyrev's mirror construction for Calabi-Yau complete intersections in toric varieties \cite{Batyrev:1993dm,Batyrev:1994pg}, one can construct a number of mirror pairs $(X,X^*)$ satisfying
\begin{equation}
\dim H^i(\wedge^jT^*X)=\dim H^i(\wedge^jTX^*),\ \ \dim H^i(\wedge^jTX)=\dim H^i(\wedge^jT^*X^*),\ \ 0\le i,j\le d,
\end{equation}
where $T^*X$ and $TX$ are holomorphic cotangent and tangent bundles on $X$, respectively. The K\"ahler moduli space ${\cal M}_{\mbox{\scriptsize K\"ahler}}(X)$ of $X$ is locally isomorphic to $H^1(\wedge^1T^*X)$, while the complex structure moduli space ${\cal M}_{\mbox{\scriptsize comp}}(X^*)$ of $X^*$ corresponds to $H^1(\wedge^1TX^*)$. Therefore the mirror symmetry implies the existence of the isomorphism between ${\cal M}_{\mbox{\scriptsize K\"ahler}}(X)$ and ${\cal M}_{\mbox{\scriptsize comp}}(X^*)$. Moreover the mirror symmetry yields the equivalence of the A-model correlators of the observables ${\cal O}^{(i)}_{\ell}$ defined by the elements $J^{(i)}_{\ell}$ in $H^i(\wedge^iT^*X)$ and the B-model correlators of the observables ${\widetilde {\cal O}}^{(i)}_{\ell}$ corresponding to $\widetilde{J}^{(i)}_{\ell}$ in $H^i(\wedge^iTX^*)$. Here the index $\ell$ just labels the elements. As a consequence of the K\"ahler moduli dependence, the A-model correlators receive $\alpha'$ corrections corresponding to the worldsheet instantons, whereas the B-model correlators do not receive these kind of corrections. Therefore the mirror symmetry provides a powerful method to compute the quantum corrections in the A-model by means of the classical calculation of the B-model.

As shown in \cite{Greene:1993vm}, the three-point A-model correlator on the worldsheet ${\IP}^1$ takes the form
\begin{equation}
C_{\ell mn}^{(i,j)}=\big<{\cal O}^{(i)}_{\ell}{\cal O}^{(j)}_{m}{\cal O}^{(d-i-j)}_{n}\big>_{\scriptsize{\IP}^1}=
\kappa_{\ell mn}^{(i,j)}+\sum_{\beta\in H_2(X,{\IZ})\setminus \{0\}}N_{\beta,\ell mn}^{(i,j)}\frac{q^{\beta}}{1-q^{\beta}},
\label{threePt}
\end{equation}
where $q^{\beta} \equiv e^{2\pi i\sum_{\ell=1}^{h^{1,1}(X)}\beta_{\ell}t^{\ell}}$ and $t^{\ell}$ are the complexified K\"ahler parameters.\footnote{This multiple cover formula of the three-point correlator is the higher dimensional analogue of the Aspinwall-Morrison formula for Calabi-Yau threefold \cite{Aspinwall:1991ce}. In the case of the Calabi-Yau threefold, $n_{\beta} \equiv \frac{N_{\beta,\ell mn}^{(1,1)}}{\beta_{\ell}\beta_m\beta_n}$ can be interpreted as the genus zero part of the Gopakumar-Vafa invariants which count the BPS states of M2 branes wrapping the holomorphic two-cycles \cite{Gopakumar:1998ii,Gopakumar:1998jq}.} The leading contributions $\kappa_{\ell mn}^{(i,j)}=\int_XJ^{(i)}_{\ell}\wedge J^{(j)}_{m}\wedge J^{(d-i-j)}_{n}$ are the classical intersection numbers and the coefficients $N_{\beta,\ell mn}^{(i,j)}$ correspond to the genus zero Gromov-Witten invariants related to the number of rational curves (the worldsheet instantons) of class $\beta$ which intersect with the cycles dual to the inserted observables. In particular, it is thought that
\begin{equation}
n_{\beta,n}=\frac{N_{\beta,\ell mn}^{(1,1)}}{\beta_{\ell}\beta_m}
\label{threeInt}
\end{equation}
are integer invariants \cite{Klemm:2007in}.\footnote{Note that these invariants are independent of the observables ${\cal O}^{(1)}_{\ell}$ and ${\cal O}^{(1)}_{m}$.} The generating function of the Gromov-Witten invariants $n_{\beta,n}$ can be defined as
\begin{equation}
F_n(t)=\frac12\sum_{\ell,m=1}^{h^{1,1}(X)}\kappa_{\ell mn}^{(1,1)}t^{\ell}t^m+\widehat{F}_{n}(t),\ \
\widehat{F}_{n}(t)=\frac{1}{(2\pi i)^2}\sum_{\beta\in H_2(X,{\IZ})\setminus \{0\}}n_{\beta,n}{\rm Li}_2(q^{\beta}),
\label{HoloF}
\end{equation}
where ${\rm Li}_k(q)=\sum_{n=1}^{\infty}\frac{q^n}{n^k}$ is the polylogarithm. Then the three-point correlation function (\ref{threePt}) with $i=j=1$ can be expressed as $C_{\ell mn}^{(1,1)}=\frac{\partial}{\partial t^{\ell}}\frac{\partial}{\partial t^m}F_n(t)$.\footnote{As discussed in \cite{Klemm:1996ts} from the perspective of mirror B-model, $C_{\ell mn}^{(1,1)}$ provides all the three-point correlators $C_{\ell mn}^{(i,j)}$. Observables of these correlators are restricted to take values at the primary subspace of the vertical cohomology $\oplus_iH^{i,i}(X)$ which is generated by wedge products of the elements of $H^{1,1}(X)$ \cite{Greene:1993vm}.} Here we have used an abridged notation $t=\{t^{\ell}\}_{\ell=1}^{h^{1,1}}$ for the complexified K\"ahler moduli.

By using the mirror symmetry, the holomorphic functions $F_n(t)$ at the large radius point $q=0$ can be determined from solutions to Picard-Fuchs equations which govern the periods of the holomorphic $d$-form $\Omega(z)$ on $X^*$. Here $z$ indicates the set of the complex structure moduli $z=\{z_{\ell}\}_{\ell=1}^{h^{d-1,1}}$. These equations can be derived from the theory of variation of the Hodge structures \cite{Ceresole:1992su,Greene:1993vm} and the solutions to the equations reflect the structure of primary subspace of the horizontal cohomology $\oplus_iH^{d-i,i}(X^*)$ generated by the cup products of the elements in $H^{d-1,1}(X^*)$ \cite{Greene:1993vm}. At the large complex structure point $z=0$, there are $h_{\mbox{\scriptsize prim}}^{d-i,i}(X^*)$ $(0<i<d)$ linearly independent solutions $\Pi_{\ell}^{(i)}(z)$ whose leading contributions are the $i$-th power of the logarithm of $z$. Then the flat coordinates of the complex structure moduli space ${\cal M}_{\mbox{\scriptsize comp}}(X^*)$ identified with the complexified K\"ahler moduli of $X$ are determined by the mirror map
\begin{equation}
t^{\ell}=\frac{\Pi_{\ell}^{(1)}(z)}{2\pi i\Pi^{(0)}(z)}\sim
\frac{1}{2\pi i}\log z_{\ell},
\end{equation}
where $\Pi^{(0)}(z)=1+{\cal O}(z)$ is a unique single-valued solution. Through the use of the mirror map, the holomorphic function (\ref{HoloF}) corresponding to the observable ${\cal O}^{(d-2)}_{n}$ restricted to the primary subspace of $H^{d-2,d-2}(X)$ can be derived from a period as
\begin{equation}
F_n(t)=\frac{\Pi_{n}^{(2)}(z)}{(2\pi i)^2\Pi^{(0)}(z)}\sim
\frac{1}{2(2\pi i)^2}\sum_{\ell,m=1}^{h^{d-1,1}(X^*)}\kappa_{\ell mn}^{(1,1)}\log z_{\ell}\log z_m,
\label{PFsecond}
\end{equation}
where $\kappa_{\ell mn}^{(1,1)}$ are identified with the classical intersection numbers of $X$.

For Calabi-Yau threefold, both the K\"ahler and the complex structure moduli spaces are the local special K\"ahler manifold \cite{Dixon:1989fj,Strominger:1990pd,Candelas:1990pi,Craps:1997gp}. The K\"ahler potential of such a manifold can be determined by the prepotential ${\cal F}(T)$ satisfying the scaling property ${\cal F}(\lambda T)=\lambda^2{\cal F}(T)$, where $T$ is a set of the special coordinates of the moduli space defined by $T=\{T^0,T^{\ell}\} \equiv \{\Pi^{(0)},\Pi^{(0)}t^{\ell}\}$. At the large complex structure point, the log-squared solutions $\Pi_{\ell}^{(2)}(z)$ correspond to ${\cal F}_{\ell}(T) \equiv \frac{\partial}{\partial T^{\ell}}{\cal F}(T)$, and a log-cubed solution $\Pi^{(3)}(z)\sim
-\frac{1}{3!(2\pi i)^3}\sum_{\ell,m}\kappa_{\ell mn}^{(1,1)}\log z_{\ell}\log z_m\log z_n$ becomes equivalent to ${\cal F}_0(T) \equiv \frac{\partial}{\partial T^0}{\cal F}(T)$. The solutions $\{ {\cal F}_0, {\cal F}_{\ell} \}$ are also known as the conjugate special coordinates of the moduli space. Using the scaling property of the prepotential, one can define ${\cal F}(T)=\big(T^0\big)^2F(t)$ and the conjugate coordinates can be expressed in terms of $t^{\ell}$ as
\begin{equation}
{\cal F}_{\ell}(T)=T^0\frac{\partial}{\partial t^{\ell}}F(t),\ \
{\cal F}_0(T)=T^0\bigg[2F(t)-\sum_{\ell=1}^{h^{1,1}(X)}t^{\ell}\frac{\partial}{\partial t^{\ell}}F(t)\bigg].
\label{PreFs}
\end{equation}
Up to linear and quadratic terms of $t^{\ell}$, the prepotential takes the form \cite{Hosono:1994ax}
\begin{align}
F(t)=\frac{1}{3!}\sum_{\ell,m,n}^{h^{1,1}(X)}\kappa_{\ell mn}t^{\ell}t^mt^n-\frac{i}{16\pi^3}\zeta(3)\chi(X)+\widehat{F}(t),
\end{align}
where
\begin{align}
\widehat{F}(t)=\frac{1}{(2\pi i)^3}\sum_{\beta\in H_2(X,{\IZ})\setminus \{0\}}n_{\beta}{\rm Li}_3(q^{\beta}).
\label{Fhat}
\end{align}
Here $\chi(X)$ is the Euler characteristic of $X$ and the integers $n_{\beta}$ are the genus zero Gromov-Witten invariants. In terms of the prepotential, the generating function (\ref{HoloF}) is expressed as $F_{\ell}(t)=\frac{\partial}{\partial t^{\ell}}F(t)$.

\subsection{K\"ahler potential and ${\cal N}=(2,2)$ GLSM}

In this subsection, we first explain the relation between the solutions to the Picard-Fuchs equations and the K\"ahler potential for Calabi-Yau threefold. Then we briefly summarize the result of \cite{Jockers:2012dk} which provides a novel prescription to compute the K\"ahler potential.

Both of the K\"ahler and the complex structure moduli spaces of Calabi-Yau $d$-fold are K\"ahler manifold equipped with the K\"ahler potential. According to the mirror symmetry, the quantum-corrected K\"ahler potential on the K\"ahler moduli space of a Calabi-Yau $d$-fold $X$ can be identified with the K\"ahler potential on the complex structure moduli space of the mirror manifold $X^*$. For Calabi-Yau threefold, the K\"ahler potential can be expressed in terms of the periods as \cite{Ferrara:1989vp,Strominger:1990pd,Candelas:1990pi}
\begin{eqnarray}
K(z,{\overline z})&=&
-\log i\int_{X^*}\Omega(z)\wedge \overline{\Omega (z)}\nonumber\\
&=&
-\log i\sum_{I}\big(\overline{T}^{I}{\cal F}_{I}(T)-T^{I}\overline{{\cal F}_{I}(T)}\big), \ \ I=0,1, \ldots, h^{1,1}(X)
\label{3KahlF}
\end{eqnarray}
up to the K\"ahler transformations $K(z,{\overline z})\to K(z,{\overline z})+f(z)+\overline{f(z)}$ where $f(z)$ is a local holomorphic function. Substituting (\ref{PreFs}) -- (\ref{Fhat}), at the large radius point we obtain
\begin{eqnarray}
e^{-K(z,{\overline{z}})}&=&
-\frac{i}{6}\sum_{\ell,m,n}\kappa_{\ell mn}(t^{\ell}-{\overline t}^{\ell})(t^m-{\overline t}^m)(t^n-{\overline t}^n)+\frac{1}{4\pi^3}\zeta(3)\chi(X)\nonumber\\
&&\hspace{-1em}
-\frac{i}{(2\pi i)^2}\sum_{\beta,\ell}n_{\beta}\big({\rm Li}_2(q^{\beta})+{\rm Li}_2({\overline q}^{\beta})\big)\beta_{\ell}(t^{\ell}-{\overline t}^{\ell})+\frac{2i}{(2\pi i)^3}\sum_{\beta}n_{\beta}\big({\rm Li}_3(q^{\beta})+{\rm Li}_3({\overline q}^{\beta})\big),\nonumber\\
&&
\label{3dKahl}
\end{eqnarray}
where we used the degrees of freedom of K\"ahler transformations to subtract $\log T^0(z)\overline{T^0(z)}$ from $K(z,{\overline{z}})$. This is a standard way to evaluate the K\"ahler potential by using the mirror symmetry. Even if we consider the cases in which no mirror construction is known, it is expected that the formula (\ref{3dKahl}) still holds.

Alternatively, as conjectured in \cite{Jockers:2012dk} and proved physically in \cite{Gomis:2012wy}, the K\"ahler potential on the K\"ahler moduli space of $X$ can be computed from the partition function of an ${\cal N}=(2,2)$ GLSM on two-sphere $S^2$ via
\begin{equation}
e^{-K(z,{\overline z})}=Z_{\mbox{\scriptsize GLSM}}.
\label{JKLMR2}
\end{equation}
Here it is assumed that the GLSM description which flows in the IR to the non-linear sigma model on $X$ exists. The Fayet-Iliopoulos (FI) parameters $r_{\ell}$ and the theta angles $\theta_{\ell}$ of the GLSM are related to the quantum K\"ahler moduli $z=\{z_{\ell}\}$ on $X$ via
\begin{equation}
z_{\ell}=e^{-2\pi r_{\ell}+i\theta_{\ell}}.
\label{GLSMKahl}
\end{equation}

The two sphere partition function $Z_{\mbox{\scriptsize GLSM}}$ was exactly computed in \cite{Benini:2012ui,Doroud:2012xw}. Suppose the gauge group is $G \times U(1)^s$. Throughout this paper we focus on the cases of $G=U(k)$. The matter sector consists of chiral multiplets $\Phi_A$ in the irreducible representation $R_A$ of $G$ and we denote their charges under the $U(1)$ factors as $Q^{\ell}_A, \ \ell=1, \ldots, s$. By performing the supersymmetric localization, the authors of \cite{Benini:2012ui,Doroud:2012xw} obtained\footnote{Here we consider the Coulomb branch representation in which we take the coefficient of the deformation term to be infinity.}
\begin{equation}
Z_{\mbox{\scriptsize GLSM}}=\frac{1}{|{\cal W}|}\sum_{\mathfrak{m}=\{ \mathfrak{m}_i, \widetilde{\mathfrak{m}}_{\ell}\} }\int\bigg[\prod_{i=1}^{\textrm{rank}(G)}\prod_{\ell=1}^{s}\frac{d\sigma_{i}}{2\pi}\frac{d\widetilde{\sigma}_{\ell}}{2\pi}\bigg]Z_{\mbox{\scriptsize class}}(\sigma, \mathfrak{m})Z_{\mbox{\scriptsize gauge}}(\{\sigma_i \} , \{ \mathfrak{m}_i \})
\prod_A Z_{\Phi_A}(\sigma, \mathfrak{m}),
\label{GLSMp}
\end{equation}
where $|{\cal W}|$ is the order of the Weyl group of $G$ and $\sigma = \{ \sigma_i, \widetilde{\sigma}_{\ell} \}$. $\sigma_{i} \in {\mathbb{R}}^{{\textrm{rank}}(G)}$ is in the Cartan subalgebra of $G$ and $\mathfrak{m}_{i} \in {\mathbb{Z}}^{{\textrm{rank}}(G)}$ is the magnetic charge for the Cartan part of the gauge group $G$ (GNO charge \cite{Goddard:1976qe}). Similarly, $\widetilde{\sigma}_{\ell}$ and $\widetilde{\mathfrak{m}}_{\ell}$ parametrize ${\mathbb{R}}^s$ and ${\mathbb{Z}}^s$ respectively.
 
The partition function consists of three pieces. The first of these is the contribution from the classical action on the localization configuration which takes the form
\begin{equation}
Z_{\mbox{\scriptsize class}}(\sigma,\mathfrak{m})=e^{-4\pi i(r\sum_i\sigma_i+\sum_{\ell}r_{\ell}\widetilde{\sigma}_{\ell})
-i(\theta\sum_i\mathfrak{m}_i+\sum_{\ell}\theta_{\ell}\widetilde{\mathfrak{m}}_{\ell})},
\label{GLSMcla}
\end{equation}
where $r$ and $\theta$ are the FI parameter and the theta angle for central $U(1) \subset G$, respectively. $Z_{\mbox{\scriptsize gauge}}(\{ \sigma_i \}, \{ \mathfrak{m}_i \})$ and $Z_{\Phi_A}(\sigma,\mathfrak{m})$ are the one loop determinants of the vector multiplet and the chiral multiplets given by
\begin{eqnarray}
\label{GLSMgau}
&&
Z_{\mbox{\scriptsize gauge}}(\{ \sigma_i \}, \{ \mathfrak{m}_i \})
=\prod_{\alpha\in\Delta_{+}}\Big(\frac{(\alpha,\mathfrak{m})^2}{4}+(\alpha,\sigma)^2\Big),\\
&&
Z_{\Phi_A}(\sigma,\mathfrak{m})
=\prod_{w\in R_{A}}\frac{\Gamma\big(\frac12\mathfrak{q}_{A}-(w,i\sigma+\frac12\mathfrak{m})
-\sum_{\ell}Q^{\ell}_A(i\widetilde{\sigma}_{\ell}+\frac12\widetilde{\mathfrak{m}}_{\ell})\big)}{\Gamma\big(1-\frac12\mathfrak{q}_{A}+(w,i\sigma-\frac12\mathfrak{m})
+\sum_{\ell}Q^{\ell}_A(i\widetilde{\sigma}_{\ell}-\frac12\widetilde{\mathfrak{m}}_{\ell})\big)},
\label{GLSMmatt}
\end{eqnarray}
where $(\cdot,\cdot)$ is the standard inner product, $\Delta_{+}$ is the set of positive roots, $w$ is the weight vector of the representation $R_{A}$, and $\mathfrak{q}_{A}$ is the $U(1)_V$ R-charge of a chiral multiplet.\footnote{Note that, as discussed in \cite{Doroud:2012xw}, the theory on $S^2$ breaks the classical $U(1)_A$ R-symmetry of the original ${\cal N}=(2,2)$ GLSM in flat space. It was also emphasized that unitarity constraints the R-charges to be non-negative. If we consider non-compact Calabi-Yau variety as target space, all the R-charges of chiral fields of GLSM should be set to zero \cite{Park:2012nn}. We will encounter these cases in Section 5.}

In the subsequent sections, we mainly focus on $d=4$ and conjecture the exact formula of the quantum-corrected K\"ahler potential for Calabi-Yau fourfold. Moreover, by utilizing the proposal (\ref{JKLMR2}), we test the conjecture by computing the Gromov-Witten invariants of Calabi-Yau fourfolds and comparing them with the existing results in the literature.

\section{K\"ahler potential for Calabi-Yau fourfolds}

Now we turn to study the K\"ahler moduli space of a Calabi-Yau fourfold $X$. For $d=4$, the non-trivial three-point A-model correlator in (\ref{threePt}) is given by
\begin{align}
C_{k \ell n}^{(1,1)}=\big<{\cal O}^{(1)}_{k}{\cal O}^{(1)}_{\ell}{\cal O}^{(2)}_{n}\big>_{\scriptsize{\IP}^1}.
\label{correl1124}
\end{align}
Let us consider the observable ${\cal O}^{(2)}_{n}$ defined on the primary subspace of the cohomology $H^{2,2}_{\mbox{\scriptsize prim}}(X) \subset H^{2,2}(X)$ generated by the wedge products $J_k \wedge J_{\ell}$, where $J_k$ are the elements of $H^{1,1}(X)$. Let $\{H_n\}$ be a basis of $H^{2,2}_{\mbox{\scriptsize prim}}(X)$. As in (\ref{HoloF}), we can define the generating function of the Gromov-Witten invariants associated with an element $H_n$ as
\begin{equation}
F_n(t)=\frac12\kappa_{k\ell n}t^{k}t^{\ell}+\widehat{F}_{n}(t),\ \
\widehat{F}_{n}(t)=\frac{1}{(2\pi i)^2}\sum_{\beta\in H_2(X,{\IZ})\setminus \{0\}}n_{\beta,n}{\rm Li}_2(q^{\beta}),
\label{HolFnpr}
\end{equation}
where $\kappa_{k\ell n}=\int_XJ_k\wedge J_{\ell}\wedge H_n$ are the classical intersection numbers.

Here we conjecture that in the vicinity of the large radius point the quantum-corrected K\"ahler potential for Calabi-Yau fourfold $X$ is given by\footnote{We have arrived at this formula from explicit computation for some examples (see Section 4) based on the relation (\ref{JKLMR2}). Note that there are no corrections proportional to $\zeta(2)$ and $\zeta(4)$ in the final results. It would be interesting to clarify the reason for the absence of these terms from the perspective of the non-linear sigma model \cite{Grisaru:1986px}.}
\begin{eqnarray}
e^{-K(z,{\overline{z}})}&=&
\frac{1}{4!}\kappa_{ijk\ell}(t^i-\overline{t}^i)(t^j-\overline{t}^j)(t^k-\overline{t}^k)(t^{\ell}-\overline{t}^{\ell})+\frac12\big(\widehat{G}_{k\ell}(t)+\overline{\widehat{G}_{k\ell}(t)}\big)(t^k-\overline{t}^k)(t^{\ell}-\overline{t}^{\ell})\nonumber\\
&&
-\big(\widehat{H}_{\ell}(t)-\overline{\widehat{H}_{\ell}(t)}\big)(t^{\ell}-\overline{t}^{\ell})
+\frac12\eta^{mn}\big(\widehat{F}_{mn;\ell}(t)-\overline{\widehat{F}_{mn;\ell}(t)}\big)(t^{\ell}-\overline{t}^{\ell})
\nonumber\\
&&
+\frac{i}{4\pi^3}\zeta(3)C_{\ell}(t^{\ell}-\overline{t}^{\ell})
-\frac12\eta^{mn}\big(\widehat{F}_m(t)-\overline{\widehat{F}_m(t)}\big)\big(\widehat{F}_n(t)-\overline{\widehat{F}_n(t)}\big)
\label{Kahl4}
\end{eqnarray}
up to the degrees of freedom of K\"ahler transformation. Here we defined the generating functions associated with the elements $J_k\wedge J_{\ell}$ as
\begin{equation}
G_{k\ell}(t)=\frac12\kappa_{ijk\ell}t^it^j+\widehat{G}_{k\ell}(t),\ \ \widehat{G}_{k\ell}(t)=\frac{1}{(2\pi i)^2}\sum_{\beta\in H_2(X,{\IZ})\setminus \{0\}}n_{\beta,k\ell}{\rm Li}_2(q^{\beta}),
\label{HolGij}
\end{equation}
where the coefficients $\kappa_{ijk\ell}=\int_XJ_i\wedge J_j\wedge J_k\wedge J_{\ell}$ are the classical quadruple intersection numbers. Note that such a generating function $G_{k\ell}(t)$ can be obtained by a linear combination of $F_n(t)$.\footnote{In general we have multiple elements of $H^{2,2}_{\mbox{\scriptsize prim}}(X)$ within the wedge product $J_k \wedge J_{\ell}$.} We have also defined
\begin{eqnarray}
\label{Hldef}
&&
\widehat{H}_{\ell}(t)=\int^{t^{\ell}}_{i\infty}\widehat{G}_{\ell\ell}(t)d{\widetilde t}^{\ell}+2\sum_{k\neq \ell}\int^{t^k}_{i\infty}\widehat{G}_{k\ell}(t)d{\widetilde t}^k\bigg|_{\scriptsize t^{\ell}=i\infty},\\
\label{Fmndef}
&&
\widehat{F}_{mn;\ell}(t)=\int^{t^{\ell}}_{i\infty}{\widetilde \partial}_{\ell}\widehat{F}_m(t){\widetilde \partial}_{\ell}\widehat{F}_n(t)d{\widetilde t}^{\ell},\\
&&
C_{\ell}=\int_{X}c_3(X)\wedge J_{\ell},
\label{topc3j}
\end{eqnarray}
and $\eta^{mn}=\eta^{-1}_{mn}$ is the inverse matrix of the intersection matrix $\eta_{mn}=\int_{X}H_m\wedge H_n$ on $H^{2,2}_{\mbox{\scriptsize prim}}(X)$. $c_3(X)$ is the third Chern class of $X$. In (\ref{Hldef}) and (\ref{Fmndef}), we abbreviated the arguments of the integrands as $t=\{t^1,\ldots,t^{\ell-1},\widetilde{t}^{\ell},t^{\ell+1},\ldots,t^{h^{1,1}}\}$. Our expression (\ref{Kahl4}) is exactly the four dimensional extension of (\ref{3dKahl}) which has not been fully understood. 

Let us revisit our conjecture (\ref{Kahl4}) from the viewpoint of the B-model.
Suppose we consider a Calabi-Yau fourfold $X$ whose mirror construction is known. As explained in Section 2, on the B-model side, the solutions to Picard-Fuchs equations give the periods of a holomorphic $d$-form $\Omega(z)$ on the mirror manifold. Here we assume that the periods take the following forms
\begin{align}
&
\Pi^{(0)}(z)=T^0(z),\ \ \frac{\Pi_{\ell}^{(1)}(z)}{2\pi iT^0(z)}=t^{\ell},\ \
\frac{\Pi_n^{(2)}(z)}{(2\pi i)^2T^0(z)}=\frac12\kappa_{\ell mn}t^{\ell}t^m+\widehat{F}_{n}(t),\nonumber \\
&
\frac{\Pi_{\ell}^{(3)}(z)}{(2\pi i)^3T^0(z)}=\frac{1}{3!}\kappa_{ijk\ell}t^it^jt^k+\widehat{G}_{k\ell}(t)t^k-\widehat{H}_{\ell}(t)+\frac12\eta^{mn}\widehat{F}_{mn;\ell}(t)+\frac{i}{8\pi^3}\zeta(3)C_{\ell},\nonumber\\
&
\frac{\Pi^{(4)}(z)}{(2\pi i)^4T^0(z)}=\frac{1}{4!}\kappa_{ijk\ell}t^it^jt^kt^{\ell}+\frac12\widehat{G}_{k\ell}(t)t^kt^{\ell}-\widehat{H}_{\ell}(t)t^{\ell}+\frac12\eta^{mn}\widehat{F}_{mn;\ell}(t)t^{\ell}\nonumber\\
&\hspace{6em}+\frac{i}{8\pi^3}\zeta(3)C_{\ell}t^{\ell}-\frac12\eta^{mn}\widehat{F}_m(t)\widehat{F}_n(t).
\label{PF4c}
\end{align}
in the vicinity of the large complex structure point.\footnote{This should be obtained by the Frobenius method as in the case of the complete intersection Calabi-Yau threefolds in projective spaces \cite{Hosono:1993qy,Hosono:1994ax} (see also \cite{Grimm:2009ef} which proposed an alternative method for Calabi-Yau fourfolds by using the analytic continuation to a conifold point and a monodromy analysis). In the context of open mirror symmetry, the relative periods of Calabi-Yau threefolds with branes were studied in  \cite{Alim:2009bx,Alim:2011rp} and it was also mentioned that these are related to the periods of Calabi-Yau fourfolds without branes. It would be interesting to provide further details of the relationship to our conjecture.} Plugging these periods into a four dimensional analogue of (\ref{3KahlF}) given by
\begin{equation}
e^{-K(z,{\overline{z}})}=\Big[\Pi^{(0)}(z){\overline{\Pi^{(4)}(z)}}+\sum_{\ell}\Pi_{\ell}^{(1)}(z){\overline{\Pi_{\ell}^{(3)}(z)}}+c.c.\Big]+\eta^{mn}\Pi_m^{(2)}(z){\overline{\Pi_n^{(2)}(z)}},
\end{equation}
we can obtain the conjectural formula (\ref{Kahl4}).

\section{Examples}

According to the relation (\ref{JKLMR2}) proposed in \cite{Jockers:2012dk}, it should be possible to verify our conjecture (\ref{Kahl4}) about the exact K\"ahler potential for Calabi-Yau fourfold by comparing with the ${\cal N}=(2,2)$ GLSM partition function. In this section, through the use of (\ref{Kahl4}), we extract the topological invariants such as the genus zero Gromov-Witten invariants from the GLSM partition function, and show that our conjecture (\ref{Kahl4}) is consistent with the mirror symmetry predictions. 

Here we explain a prescription to extract the topological data from GLSM calculation. As we mentioned in Section 2.2, the quantum K\"ahler moduli $z_{\ell}$ on a Calabi-Yau manifold $X$ are related to the FI parameters and the theta angles of the corresponding GLSM. In order to evaluate the K\"ahler potential in the vicinity of the large radius point $z_{\ell}=0$, we need to find the flat coordinates $t^{\ell}$ which give the classical K\"ahler moduli. As explained in \cite{Jockers:2012dk}, the flat coordinates can be determined by the following procedure. First we perform the contour integration of the two sphere partition function $Z_{\mbox{\scriptsize GLSM}}$ around the large radius point. As indicated in (\ref{Kahl4}), the coefficients of $\frac{1}{4!} \log \overline{z}_i\log \overline{z}_j\log \overline{z}_k\log \overline{z}_{\ell}$ in $Z_{\mbox{\scriptsize GLSM}}$ should be the classical intersection numbers $\kappa_{ijk\ell}$. Note that we need to use the degrees of freedom of K\"ahler transformation $K(z,{\overline z})\to K(z,{\overline z})+f(z)+\overline{f(z)}$ in order to obtain appropriate intersection numbers. This corresponds to the normalization of the partition function.\footnote{As we will see later, $f(z)$ coincides with the logarithm of a solution $\Pi^{(0)}(z)=T^0(z)$ to the corresponding Picard-Fuchs equations.} This is similar to the situation in the Calabi-Yau threefold in \cite{Jockers:2012dk}. After performing the K\"ahler transformation, from the coefficients of $\log \overline{z}_i\log \overline{z}_j\log \overline{z}_k$ which should be identified with
$\frac{1}{3! (2\pi i)^3}\kappa_{ijk\ell}t^{\ell}$, we can determine the flat coordinates which take the form
\begin{equation}
2\pi it^{\ell}=\log z_{\ell}+2\pi it^{\ell}_{(0)}+\Delta_{\ell}(z).
\label{flatGLSM}
\end{equation}
Here $\Delta_{\ell}(z)$ are holomorphic functions and $0\le t^{\ell}_{(0)}<1$ are constants fixed by requiring the positivity of Gromov-Witten invariants. By inverting (\ref{flatGLSM}), we can express the $z_{\ell}$ in terms of $t^{\ell}$ as $z_{\ell}=e^{-2\pi i t^{\ell}_{(0)}}q_{\ell}+{\cal O}(q^2)$, where $q_{\ell}=e^{2\pi i t^{\ell}}$. Reading off the coefficients of $\log \overline{z}_k\log \overline{z}_{\ell}$ in the $q$-expansion, we obtain the generating functions (\ref{HolGij}).\footnote{Let $r$ be the number of the independent elements of $\{J_k\wedge J_{\ell}\}$. In cases where $\dim H^{2,2}_{\mbox{\scriptsize prim}}(X)$ equals to $r$, the above prescription determines all the generating function (\ref{HolFnpr}). In cases of $\dim H^{2,2}_{\mbox{\scriptsize prim}}(X)=r+1$, as in Section 4.4 we can extract one remaining generating function in (\ref{HolFnpr}) by fixing the intersection matrix $\eta_{mn}$. However in cases of $\dim H^{2,2}_{\mbox{\scriptsize prim}}(X)= r+s,\ s\ge 2$, the remaining $s$ generating functions in (\ref{HolFnpr}) can not be determined.}

\subsection{Sextic fourfold: $X_6\subset {\IP}^5$}

As a simplest example of the compact Calabi-Yau fourfold, let us consider the Fermat sextic fourfold $X_6\subset {\IP}^5$ defined by a degree six hypersurface in ${\IP}^5$ \cite{Klemm:2007in} (see also \cite{Grimm:2009ef}). There is one K\"ahler modulus associated with the radius of ${\IP}^5$. First we determine the topological data of this manifold. Using the K\"ahler form $J$ of ${\IP}^5$, the classical quadruple intersection number is computed as $\kappa=\int_{X_6}J^4=\int_{{\IP}^5}6J^5=6$. From the total Chern class $c(X_6)=\frac{(1+J)^6}{1+6J}$, we can also see that $\int_{X_6}c_3(X_6)\wedge J=-70\int_{X_6}J^3 \wedge J=-420$.

\begin{table}[t]
\begin{center}
\begin{tabular}{c|c|c}
\hline
Field & U(1) & $U(1)_V$ \\ \hline
$\Phi_i$ & +1 & $2\mathfrak{q}$ \\
$P$ & $-6$ & $2-12\mathfrak{q}$ \\ \hline
\end{tabular}
\caption{Matter content of the abelian GLSM for the sextic fourfold in ${\mathbb{P}}^5$. Here $i=1, \ldots, 6$. We set the R-charges in such a way that the total R-charge of superpotential becomes $2$. The positivity of the R-charges implies $0<\mathfrak{q}<\frac16$.}
\label{sextic}
\end{center}
\end{table}

The sextic fourfold has an abelian ${\cal N}=(2,2)$ GLSM description with matter content shown in Table \ref{sextic}. This model has a superpotential $W=P W_6(\Phi)$ where $W_6(\Phi)$ is a homogeneous degree six polynomial of $\Phi_i$. The GLSM has a phase transition which occurs as the FI parameter $r$ is varied. Here we consider the Calabi-Yau phase $r \gg 0$.

Using the formulas (\ref{GLSMp}) -- (\ref{GLSMmatt}), we can write the exact GLSM partition function for the sextic fourfold as
\begin{eqnarray}
Z_{\mbox{\scriptsize GLSM}}&=&
\sum_{m\in{\IZ}}e^{-i\theta m}\int_{-\infty}^{\infty}\frac{d\sigma}{2\pi}e^{-4\pi ir\sigma}\frac{\Gamma(\mathfrak{q}-i\sigma-\frac{1}{2}m)^6}{\Gamma(1-\mathfrak{q}+i\sigma-\frac{1}{2}m)^6}\frac{\Gamma(1-6\mathfrak{q}+6i\sigma+3m)}{\Gamma(6\mathfrak{q}-6i\sigma+3m)}.
\end{eqnarray}
This can be evaluated in the same way as performed in \cite{Jockers:2012dk} and we obtain
\begin{eqnarray}
Z_{\mbox{\scriptsize GLSM}}
&=&
(z\overline{z})^\mathfrak{q}\oint\frac{d\epsilon}{2\pi i}(z\overline{z})^{-\epsilon}\frac{\pi^5\sin(6\pi\epsilon)}{\sin^6(\pi \epsilon)}\bigg|\sum_{k=0}^{\infty}z^k\frac{\Gamma(1+6k-6\epsilon)}{\Gamma(1+k-\epsilon)^6}\bigg|^2,
\label{sexticGLSM}
\end{eqnarray}
where $z=e^{-2\pi r+i\theta}$. Note that the complex conjugation does not act on $\epsilon$.

As explained above, we first look at the coefficient of $\log^4\overline{z}$. The result is given by
\begin{equation}
\frac{6}{4!}(z\overline{z})^\mathfrak{q}T^0(z)\overline{T^0(z)},\ \ T^0(z)=\sum_{k=0}^{\infty}z^k\frac{\Gamma(1+6k)}{\Gamma(1+k)^6}.
\end{equation}
We can show that $T^0(z)$ is a kernel of the Picard-Fuchs operator associated with the mirror manifold of the sextic fourfold given by
\begin{equation}
{\cal D}=\Theta^5-6z\prod_{k=1}^5(6\Theta+k),\ \ \Theta=z\frac{\partial}{\partial z}.
\label{PDsextic}
\end{equation}

After dividing the partition function $Z_{\mbox{\scriptsize GLSM}}$ by $(z\overline{z})^\mathfrak{q}(2\pi i)^4T^0\overline{T^0}$, we turn to determine the flat coordinate. From the coefficient of $\log^3\overline{z}$, we can read off the flat coordinate
\begin{equation}
2\pi it=\log z+2\pi it_{(0)}+\frac{6}{T^0(z)}\sum_{k=1}^{\infty}\frac{(6k)!}{(k!)^6}z^k\big[\Psi(1+6k)-\Psi(1+k)\big],
\end{equation}
where $\Psi(x)=\frac{d}{dx}\log \Gamma(x)$ is the digamma function and $t_{(0)}$ is a constant. By inverting this flat coordinate and substituting it into the partition function (\ref{sexticGLSM}), we can extract the Gromov-Witten 
invariants. With the choice of $t_{(0)}=0$, the generating function of these invariants (\ref{HolGij}) associated with $J^2$ is obtained as
\begin{eqnarray}
G(t)=\frac62t^2+\frac{1}{(2\pi i)^2}\sum_{d=1}^{\infty}n_{d}{\rm Li}_2(q^d),\ \ q=e^{2\pi i t},
\label{SextGt}
\end{eqnarray}
where the Gromov-Witten invariants $n_d$ are summarised in Table \ref{sexticGW}. The result is in perfect agreement with the mirror symmetry predictions of \cite{Greene:1993vm,Klemm:2007in}. We can also find that the explicit form of $Z_{\mbox{\scriptsize GLSM}}$ matches with our conjecture (\ref{Kahl4}) via (\ref{JKLMR2}).
\begin{table}[t]
\begin{center}
\begin{tabular}{c|r}
\hline
$d$ & $n_{d}$ \\ \hline
1 & 60480 \\
2 & 440884080  \\
3 & 6255156277440  \\
4 & 117715791990353760  \\ \hline
\end{tabular}
\caption{Gromov-Witten invariants for $X_6\subset {\IP}^5$.}
\label{sexticGW}
\end{center}
\end{table}

We can also consider general complete intersection Calabi-Yau manifolds $X_{d_1,\ldots,d_r}\subset {\IP}^n$ defined by $r$ hypersurfaces with the degrees $(d_1,\ldots,d_r)$ in the projective space ${\IP}^n$. The complex dimension of these manifolds is determined by $n-r$, and the Calabi-Yau condition is satisfied when $d_1+\cdots+d_r=n+1$. Then, including the above example, we can construct seven such fourfolds
\begin{equation}
X_6\subset {\IP}^5,~X_{2,5}\subset {\IP}^6,~X_{3,4}\subset {\IP}^6,~X_{2,2,4}\subset {\IP}^7,~X_{2,3,3}\subset {\IP}^7,~X_{2,2,2,3}\subset {\IP}^8,~X_{2,2,2,2,2}\subset {\IP}^9.
\label{CICYPn}
\end{equation}

\begin{table}[t]
\begin{center}
\begin{tabular}{c|c|c}
\hline
Field & U(1) & $U(1)_V$ \\ \hline
$\Phi_i$ & +1 & $2\mathfrak{q}$ \\
$P_a$ & $-d_a$ & $2-2d_a\mathfrak{q}$ \\ \hline
\end{tabular}
\caption{Matter content of the abelian GLSMs for the complete intersection Calabi-Yau fourfolds (\ref{CICYPn}). Here $i=1, \ldots, n+1$ and $a=1,\ldots,r$.}
\label{CICYPtable}
\end{center}
\end{table}
\noindent All these examples have one K\"ahler form $J$. The GLSM for each fourfold has the matter content shown in Table \ref{CICYPtable}. There are $r$ superpotentials $W_a=P_a W_{d_a}(\Phi),$ $a=1,\ldots,r$ where $W_{d_a}(\Phi)$ is a homogeneous degree ${d_a}$ polynomial of $\Phi_i$. In Table \ref{cicypGW} we summarize our results for Gromov-Witten invariants $n_d$ associated with $J^2$ calculated in the same way as the sextic example. We have also checked our conjecture (\ref{Kahl4}) holds in these examples. 
\begin{table}[t]
\begin{center}
\begin{tabular}{c|r|r|r}
\hline
$$ & $X_{2,5}\subset {\IP}^6\ (\kappa=10)$ & $X_{3,4}\subset {\IP}^6\ (\kappa=12)$ & $X_{2,2,4}\subset {\IP}^7\ (\kappa=16)$ \\ \hline
$n_1$ & 24500 & 16128 & 11776 \\
$n_2$ & 48263250 & 17510976 & 7677952 \\
$n_3$ & 181688069500 & 36449586432 & 9408504320 \\
$n_4$ & 905026660335000 & 100346754888576 & 15215566524416 \\
$n_5$ & 5268718476406938000 & 322836001522723584 & 28735332663693824 \\
\hline
$$ & $X_{2,3,3}\subset {\IP}^7\ (\kappa=18)$ & $X_{2,2,2,3}\subset {\IP}^8\ (\kappa=24)$ & $X_{2,2,2,2,2}\subset {\IP}^9\ (\kappa=32)$ \\ \hline
$n_1$ & 9396 & 6912 & 5120 \\
$n_2$ & 4347594 & 1919808 & 852480 \\
$n_3$ & 3794687028 & 988602624 & 259476480 \\
$n_4$ & 4368985908840 & 669909315456 & 103646279680 \\
$n_5$ & 5873711971817268 & 529707745490688 & 48276836019200 \\
\hline
\end{tabular}
\caption{Gromov-Witten invariants for complete intersection Calabi-Yau fourfolds in the projective space. Here $\kappa$ is the classical quadruple intersection number.}
\label{cicypGW}
\end{center}
\end{table}

\subsection{Quintic fibration over ${\IP}^1$: $X_{2,5}\subset{\IP}^1\times{\IP}^4$}

As an example with two K\"ahler moduli, we consider a quintic fibration over ${\IP}^1$ expressed by $X_{2,5}\subset{\IP}^1\times{\IP}^4$ which is defined as a Calabi-Yau hypersurface with degree two and degree five for the coordinates of ${\IP}^1$ and ${\IP}^4$, respectively \cite{Klemm:2007in}. Denoting the K\"ahler forms on ${\IP}^1$ and ${\IP}^4$ by $J_1$ and $J_2$, the nonzero classical quadruple intersection numbers are computed as $\kappa_{1222}=\int_{X_{2,5}}J_1\wedge J_2^3=\int_{{\IP}^1\times{\IP}^4}J_1\wedge J_2^3\wedge (2J_1+5J_2)=5$ and $\kappa_{2222}=\int_{X_{2,5}}J_2^4=\int_{{\IP}^1\times{\IP}^4}J_2^4\wedge (2J_1+5J_2)=2$. The total Chern class of this manifold is $c(X_{2,5})=\frac{(1+J_1)^2(1+J_2)^5}{1+2J_1+5J_2}$ and thus we see that $\int_{X_{2,5}}c_3(X_{2,5})\wedge J_1=-200$ and $\int_{X_{2,5}}c_3(X_{2,5})\wedge J_2=-330$.
\begin{table}[t]
\begin{center}
\begin{tabular}{c|c|c|c}
\hline
Field & $U(1)_1$ & $U(1)_2$ & $U(1)_V$ \\ \hline
$\Phi_{1,i_1}$ & +1 & 0 & $2\mathfrak{q}_1$ \\
$\Phi_{2,i_2}$ & 0 & +1 & $2\mathfrak{q}_2$ \\
$P$ & $-2$ & $-5$ & $2-4\mathfrak{q}_1-10 \mathfrak{q}_2$ \\ \hline
\end{tabular}
\caption{Matter content of the $U(1)_1 \times U(1)_2$ GLSM for the Calabi-Yau $X_{2,5}\subset{\IP}^1\times{\IP}^4$. Here $i_1=1, 2$ and $i_2=1, \ldots, 5$. The R-charges are assigned so that the total R-charge of superpotential is $2$. The positivity of R-charges requires $\mathfrak{q}_1>0,~\mathfrak{q}_2>0$, and $2\mathfrak{q}_1+5\mathfrak{q}_2<1$.}
\label{quinticfib}
\end{center}
\end{table}

Corresponding ${\cal N}=(2,2)$ GLSM has two $U(1)$ gauge groups with matter fields summarised in Table \ref{quinticfib}. These fields interact through a superpotential $W=PW_{2,5}(\Phi_1,\Phi_2)$, where $W_{2,5}(\Phi_1,\Phi_2)$ is a homogeneous degree two and degree five polynomial of $\Phi_{1,i_1}$ and $\Phi_{2,i_2}$, respectively. There are two FI parameters associated with $U(1)_1 \times U(1)_2$ gauge symmetry and we consider the Calabi-Yau phase $r_1,\ r_2\gg 0$.

According to the localization formulas (\ref{GLSMp}) -- (\ref{GLSMmatt}), the exact partition function for the quintic fibration $X_{2,5}$ is given by
\begin{align}
Z_{\mbox{\scriptsize GLSM}}=&
\sum_{m_1,m_2\in{\IZ}}e^{-i(\theta_1m_1+\theta_2m_2)}\int_{-\infty}^{\infty}\frac{d\sigma_1}{2\pi}\frac{d\sigma_2}{2\pi}e^{-4\pi i(r_1\sigma_1+r_2\sigma_2)}\nonumber \\
&
\times\frac{\Gamma(\mathfrak{q}_1-i\sigma_1-\frac{1}{2}m_1)^2}{\Gamma(1-\mathfrak{q}_1+i\sigma_1-\frac{1}{2}m_1)^2}\frac{\Gamma(\mathfrak{q}_2-i\sigma_2-\frac{1}{2}m_2)^5}{\Gamma(1-\mathfrak{q}_2+i\sigma_2-\frac{1}{2}m_2)^5}\nonumber \\
&\times \frac{\Gamma\big(1-2\mathfrak{q}_1-5 \mathfrak{q}_2+i(2\sigma_1+5\sigma_2)+(m_1+\frac52m_2)\big)}{\Gamma\big(2\mathfrak{q}_1+5 \mathfrak{q}_2-i(2\sigma_1+5\sigma_2)+(m_1+\frac52m_2)\big)}.
\end{align}
In a similar manner to the sextic fourfold, we can evaluate this into the form
\begin{align}
Z_{\mbox{\scriptsize GLSM}}= \ &(z_1\overline{z}_1)^{\mathfrak{q}_1}(z_2\overline{z}_2)^{\mathfrak{q}_2}\oint\frac{d\epsilon_1}{2\pi i}\frac{d\epsilon_2}{2\pi i}(z_1\overline{z}_1)^{-\epsilon_1}(z_2\overline{z}_2)^{-\epsilon_2}
\frac{\pi^6\sin\pi(2\epsilon_1+5\epsilon_2)}{\sin^2(\pi \epsilon_1)\sin^5(\pi \epsilon_2)}\nonumber\\
&\hspace{6em}
\times\bigg|\sum_{k_1,k_2=0}^{\infty}z_1^{k_1}(-z_2)^{k_2}\frac{\Gamma(1+(2k_1+5k_2)-(2\epsilon_1+5\epsilon_2))}{\Gamma(1+k_1-\epsilon_1)^2\Gamma(1+k_2-\epsilon_2)^5}\bigg|^2,
\end{align}
where we defined $z_{\ell}=e^{-2\pi r_{\ell}+i\theta_{\ell}}$ and the complex conjugation does not act on $\epsilon_{1,2}$.

Let us consider $\log^4\overline{z}_2$ term. The coefficient is given by
\begin{align}
\frac{2}{4!}(z_1\overline{z}_1)^{\mathfrak{q}_1}(z_2\overline{z}_2)^{\mathfrak{q}_2}T^0(z_1,z_2)\overline{T^0(z_1,z_2)},
\end{align}
where
\begin{align}
T^0(z_1,z_2)=\sum_{k_1,k_2=0}^{\infty}z_1^{k_1}(-z_2)^{k_2}\frac{\Gamma(1+2k_1+5k_2)}{\Gamma(1+k_1)^2\Gamma(1+k_2)^5}.
\end{align}
We see that $T^0(z_1,-z_2)$ is annihilated by the Picard-Fuchs operators associated with $X_{2,5}$ defined by
\begin{eqnarray}
{\cal D}_1&=&\Theta_1^2-z_1\prod_{k=1}^2(2\Theta_1+5\Theta_2+k),\nonumber\\
{\cal D}_2&=&(2\Theta_1-5\Theta_2)\Theta_2^3-4z_1(2\Theta_1+5\Theta_2+1)\Theta_2^3+25z_2\prod_{k=1}^4(2\Theta_1+5\Theta_2+k),
\end{eqnarray}
where $\Theta_{\ell}=z_{\ell}\frac{\partial}{\partial z_{\ell}}$. 

Normalizing the partition function $Z_{\mbox{\scriptsize GLSM}}$ by $(z_1\overline{z}_1)^{\mathfrak{q}_1}(z_2\overline{z}_2)^{\mathfrak{q}_2}(2\pi i)^4T^0\overline{T^0}$, we can read off the flat coordinates from the coefficients of $\log^3\overline{z}_2$ and $\log \overline{z}_1\log^2\overline{z}_2$ as
\begin{eqnarray}
2\pi it^1&=&\log z_1+2\pi it_{(0)}^1\nonumber\\
&&
+\frac{2}{T^0}\sum_{k_1,k_2=0}^{\infty}\frac{(2k_1+5k_2)!}{(k_1!)^2(k_2!)^5}z_1^{k_1}(-z_2)^{k_2}\big[\Psi(1+2k_1+5k_2)-\Psi(1+k_1)\big],\\
2\pi it^2&=&\log z_2+2\pi it_{(0)}^2\nonumber\\
&&
+\frac{5}{T^0}\sum_{k_1,k_2=0}^{\infty}\frac{(2k_1+5k_2)!}{(k_1!)^2(k_2!)^5}z_1^{k_1}(-z_2)^{k_2}\big[\Psi(1+2k_1+5k_2)-\Psi(1+k_2)\big],
\end{eqnarray}
where $0\le t_{(0)}^1, t_{(0)}^2<1$ are constants. Inverting these flat coordinates and taking $t_{(0)}^1=0$, $t_{(0)}^2=\frac12$, we can finally obtain the generating functions (\ref{HolGij}) as
\begin{eqnarray}
G_{12}(t_1,t_2)&=&
\frac52t_2^2+\frac{1}{(2\pi i)^2}\mathop{\sum_{d_1,d_2=0}^{\infty}}\limits_{(d_1,d_2)\ne (0,0)}n_{d_1,d_2,12}{\rm Li}_2(q_1^{d_1}q_2^{d_2}),\\
G_{22}(t_1,t_2)&=&
5t_1t_2+\frac22t_2^2+\frac{1}{(2\pi i)^2}\mathop{\sum_{d_1,d_2=0}^{\infty}}\limits_{(d_1,d_2)\ne (0,0)}n_{d_1,d_2,22}{\rm Li}_2(q_1^{d_1}q_2^{d_2}),
\end{eqnarray}
where the Gromov-Witten invariants are listed in Table \ref{quifibGW} and we see that these integer invariants completely agree with the result of \cite{Klemm:2007in}. We have also checked that the exact partition function coincides with (\ref{Kahl4}) up to $q_1^4$ and $q_2^4$.
\begin{table}[t]
\begin{center}
\begin{tabular}{c|rrrrr}
\hline
$n_{d_1,d_2,12}$ & $d_1=0$ & 1 & 2 &  3& 4 \\ \hline
$d_2=0$ &  & 125 & 0 & 0 & 0 \\
1 & 2875 & 195875 & 1248250 & 1799250 & 662875 \\
2 & 1218500 & 369229625 & 10980854250 & 101591346500 & 384568351000 \\ \hline
$n_{d_1,d_2,22}$ & $d_1=0$ & 1 & 2 &  3& 4 \\ \hline
$d_2=0$ &  & 0 & 0 & 0 & 0 \\
1 & 9950 & 171750 & 609500 & 609500 & 171750 \\
2 & 5487450 & 533197250 & 9651689750 & 63917722000 & 188112166000 \\ \hline
\end{tabular}
\caption{Gromov-Witten invariants for $X_{2,5}\subset{\IP}^1\times{\IP}^4$. Note that the assignments of $d_1$ and $d_2$ are exchanged in \cite{Klemm:2007in}.}
\label{quifibGW}
\end{center}
\end{table}

\subsection{Resolved determinantal sextic in ${\IP}^5$: $X_A\subset{\IP}^5\times{\IP^5}$}

\begin{table}[t]
\begin{center}
\begin{tabular}{c|c|c|c}
\hline
Field & U(1) & U(1) & $U(1)_V$ \\ \hline
$\Phi_i$ & +1 & 0 & $2\mathfrak{q}_{\phi}$ \\
$P_a$ & $-1$ & +1 & $2-2\mathfrak{q}_x-2\mathfrak{q}_{\phi}$ \\
$X_a$ & 0 & $-1$ & $2\mathfrak{q}_x$ \\
\hline
\end{tabular}
\caption{Matter content of the abelian PAX GLSM for the determinantal sextic fourfold in ${\mathbb{P}}^5$. Here $i=1, \ldots, 6$ and $a=1, \ldots, 6$.}
\label{PAX1}
\end{center}
\end{table}
Here we consider the linear determinantal Calabi-Yau fourfold defined by
\begin{align}
Z(A,5)=\{ \phi \in {\mathbb{P}}^5 \ | \ \textrm{rank} (A^i\phi_i) \leq 5 \},
\label{determsix}
\end{align}
where the $A^i$ are six $5 \times 5$ constant matrices and the $\phi_i$ are the homogeneous coordinates of ${\mathbb{P}}^5$. A GLSM construction for determinantal manifolds was studied in \cite{Jockers:2012zr}. Following their prescription, we analyze the determinantal sextic fourfold (\ref{determsix}) using the $U(1) \times U(1)$ ``PAX'' model with matter content shown in Table \ref{PAX1}. These matter multiplets interact through a superpotential $W=\textrm{tr} (PA^i\Phi_i X)$. This model has three distinct geometric phases and the ``$X_A$ phase'' \cite{Jockers:2012zr,Jockers:2012dk}
\begin{align}
X_A=\{ (\phi,x) \in {\mathbb{P}}^5\times {\mathbb{P}}^5 \ | \ (A^i\phi_i)x=0 \}
\label{detXA}
\end{align}
gives a resolution of the determinantal variety (\ref{determsix}). Here we denote the K\"ahler form on the first (base) ${\IP}^5$ by $J_1$, and the hyperplane class of the second (fiber) ${\IP}^5$ by $J_2$. Then the classical quadruple intersection numbers are computed as $\kappa_{1111}=\kappa_{2222}=6$, $\kappa_{1112}=\kappa_{1222}=15$, and $\kappa_{1122}=20$ from the top Chern class of a rank six normal bundle ${\mathcal X}$ whose total Chern class is given by $c({\mathcal X})=(1+J_1+J_2)^6$. The total Chern class of $X_A$ is given by $c(X_A)=\frac{(1+J_1)^6(1+J_2)^6}{c({\mathcal X})}$, and the other topological invariants are also obtained as $\int_{X_A}c_3(X_A)\wedge J_1=\int_{X_A}c_3(X_A)\wedge J_2=-210$ \cite{Jockers:2012zr}.

The partition function of this PAX GLSM is given by 
\begin{align}
Z_{\mbox{\scriptsize GLSM}}=&
\sum_{m_1,m_2\in{\IZ}}e^{-i(\theta_1m_1+\theta_2m_2)}\int_{-\infty}^{\infty}\frac{d\sigma_1}{2\pi}\frac{d\sigma_2}{2\pi}e^{-4\pi i(r_1\sigma_1+r_2\sigma_2)}\nonumber \\
&
\times\frac{\Gamma(\mathfrak{q}_{\phi}-i\sigma_1-\frac{1}{2}m_1)^6}{\Gamma(1-\mathfrak{q}_{\phi}+i\sigma_1-\frac{1}{2}m_1)^6}
\frac{\Gamma(\mathfrak{q}_x+i\sigma_2+\frac{1}{2}m_2)^6}{\Gamma(1-\mathfrak{q}_x-i\sigma_2+\frac{1}{2}m_2)^6}\nonumber\\
&
\times\frac{\Gamma\big(1-\mathfrak{q}_{\phi}-\mathfrak{q}_x+i(\sigma_1-\sigma_2)+\frac12(m_1-m_2)\big)^6}{\Gamma\big(\mathfrak{q}_{\phi}+\mathfrak{q}_x-i(\sigma_1-\sigma_2)+\frac12(m_1-m_2)\big)^6}.
\end{align}
Here we take a phase $r_1 \gg 0, \ r_2 \ll 0$ corresponding to the $X_A$ phase, where $r_1$ and $r_2$ correspond to $J_1$ and $J_2$, respectively. As in the previous examples, the partition function can be evaluated as
\begin{align}
Z_{\mbox{\scriptsize GLSM}}=&
(z_1 \overline{z}_1)^{\mathfrak{q}_{\phi}}(z_2 \overline{z}_2)^{\mathfrak{q}_x}\oint \frac{d \epsilon_1}{2\pi i}\frac{d \epsilon_2}{2\pi i}(z_1 \overline{z}_1)^{-\epsilon_1}(z_2 \overline{z}_2)^{-\epsilon_2}\frac{\pi^6 \sin^6 \pi (\epsilon_1+\epsilon_2)}{\sin^6{(\pi \epsilon_1)}\sin^6{(\pi \epsilon_2)}} \nonumber \\
& \times \Big| \sum_{k_1, k_2=0}^{\infty}z_1^{k_1}z_2^{k_2} \frac{\Gamma (1+k_1+k_2-\epsilon_1-\epsilon_2)^6 }{\Gamma (1+k_1-\epsilon_1)^6 \Gamma (1+k_2-\epsilon_2)^6} \Big|^2,
\label{XApp}
\end{align}
where $z_1=e^{-2 \pi r_1 +i\theta_1}$ and $z_2=e^{2 \pi r_2 -i\theta_2}$.\footnote{Also in the ``$X_{A^T}$ phase'' \cite{Jockers:2012zr} corresponding to $r_1+r_2 \gg 0, \ r_2 \gg 0$, we see that the GLSM partition function takes the same form (\ref{XApp}) with $z_1=e^{-2 \pi (r_1+r_2) +i(\theta_1+\theta_2)}$ and $z_2=e^{-2 \pi r_2 +i\theta_2}$.}

The coefficient of $\log^4{\overline{z}_1}$ term is
\begin{align}
\frac{6}{4!}(z_1 \overline{z}_1)^{\mathfrak{q}_{\phi}}(z_2 \overline{z}_2)^{\mathfrak{q}_x}T^0(z_1,z_2)\overline{
T^0(z_1,z_2)},
\end{align}
where
\begin{align}
T^0(z_1,z_2)=\sum_{k_1, k_2=0}^{\infty}z_1^{k_1}z_2^{k_2} \frac{\Gamma (1+k_1+k_2)^6 }{\Gamma (1+k_1)^6 \Gamma (1+k_2)^6}.
\label{detp0}
\end{align}
As before, after normalizing the partition function by $(z_1\overline{z}_1)^{\mathfrak{q}_{\phi}}(z_2\overline{z}_2)^{\mathfrak{q}_x}(2\pi i)^4T^0\overline{T^0}$, from the coefficients of $\log \overline{z}_1\log^2\overline{z}_2$ and $\log^3\overline{z}_2$, the flat coordinates are determined as
\begin{align}
&
2\pi it^1=\log z_1+2\pi it_{(0)}^1
+\Delta(z_1,z_2),\ \ 2\pi it^2=\log z_2+2\pi it_{(0)}^2
+\Delta(z_2,z_1),\nonumber\\
&
\Delta(z_1,z_2)=\frac{6}{T^0}\sum_{k_1,k_2=0}^{\infty}\frac{((k_1+k_2)!)^6}{(k_1!)^6(k_2!)^6}z_1^{k_1}z_2^{k_2}\big[\Psi(1+k_1+k_2)-\Psi(1+k_1)\big],
\label{detp1}
\end{align}
where $0\le t_{(0)}^1, t_{(0)}^2<1$ are constants. By taking $t_{(0)}^1=t_{(0)}^2=0$, we obtain three independent generating functions (\ref{HolGij}) associated with $J_1^2$, $J_1\wedge J_2$, and $J_2^2$ as
\begin{eqnarray}
G_{11}(t_1,t_2)&=&
\frac62t_1^2+15t_1t_2+\frac{20}{2}t_2^2+\frac{1}{(2\pi i)^2}\mathop{\sum_{d_1,d_2=0}^{\infty}}\limits_{(d_1,d_2)\ne (0,0)}n_{d_1,d_2,11}{\rm Li}_2(q_1^{d_1}q_2^{d_2}),\\
G_{12}(t_1,t_2)&=&
\frac{15}{2}t_1^2+20t_1t_2+\frac{15}{2}t_2^2+\frac{1}{(2\pi i)^2}\mathop{\sum_{d_1,d_2=0}^{\infty}}\limits_{(d_1,d_2)\ne (0,0)}n_{d_1,d_2,12}{\rm Li}_2(q_1^{d_1}q_2^{d_2}),
\label{detG12}
\end{eqnarray}
and $G_{22}(t_1,t_2)=G_{11}(t_2,t_1)$, where the Gromov-Witten invariants are listed in Table \ref{detrmsexGW}. We have also checked the conjecture (\ref{Kahl4}) is in agreement with the GLSM partition function. The sextic fourfold discussed in Section 4.1 is obtained from this example by an extremal transition. Then $G_{11}(t,0)$ should coincide with $G(t)$ in (\ref{SextGt}) \cite{Li:1998hba}.\footnote{We would like to thank the referee for a remark on this relation.} We have checked this coincidence up to $d_1=4$.
\begin{table}[t]
\begin{center}
\begin{tabular}{c|rrrrr}
\hline
$n_{d_1,d_2,11}$ & $d_1=0$ & 1 & 2 & 3 & 4 \\ \hline
$d_2=0$ &  & 210 & 0 & 0 & 0 \\
1 & 0 & 5670 & 59430 & 100170 & 34650 \\
2 & 0 & 24360 & 2579640 & 47382930 & 264433680 \\
3 & 0 & 24360 & 28015260 & 2324403900 & 55841697870 \\
4 & 0 & 5670 & 107096220 & 38404166850 & 2848564316640 \\ 
5 & 0 & 210 & 165382980 & 277070715810 & 60035324018880 \\
\hline
$n_{d_1,d_2,12}$ & $d_1=0$ & 1 & 2 & 3 & 4 \\ \hline
$d_2=0$ &  & 105 & 0 & 0 & 0 \\
1 & 105 & 6930 & 50715 & 71085 & 21420 \\
2 & 0 & 50715 & 3166800 & 45928155 & 221593050 \\
3 & 0 & 71085 & 45928155 & 2851172100 & 57546197940 \\
4 & 0 & 21420 & 221593050 & 57546197940 & 3492450469200 \\
5 & 0 & 945 & 413457450 & 493317415605 & 85788539294850 \\
\hline
\end{tabular}
\caption{Gromov-Witten invariants for $X_A$.}
\label{detrmsexGW}
\end{center}
\end{table}

By using the holomorphic function (\ref{detp0}) which should be identified with the fundamental period of the mirror manifold of $X_A$, we can find the Picard-Fuchs operators
\begin{eqnarray}
{\cal{D}}_1&=&
(\Theta_1+\Theta_2)(5\Theta_1^2-8\Theta_1\Theta_2+5\Theta_2^2)-z_1 (5\Theta_1^3+27\Theta_1^2\Theta_2+54\Theta_1\Theta_2^2+42\Theta_2^3+15\Theta_1^2 \nonumber \\
&&
+54\Theta_1\Theta_2+54\Theta_2^2+15\Theta_1+27\Theta_2+5)-z_2(42\Theta_1^3+54\Theta_1^2\Theta_2+27\Theta_1\Theta_2^2+5\Theta_2^3 \nonumber \\
\label{PFdet1}
&&
+54\Theta_1^2+54\Theta_1\Theta_2+15\Theta_2^2+27\Theta_1+15\Theta_2+5),\\
{\cal{D}}_2&=& (\Theta_1-\Theta_2)(5\Theta_1^2-4\Theta_1\Theta_2+5\Theta_2^2)-z_1 (\Theta_1+\Theta_2+1)(5\Theta_1^2+16\Theta_1\Theta_2+14\Theta_2^2+10\Theta_1 \nonumber \\
&&
+16\Theta_2+5)+z_2 (\Theta_1+\Theta_2+1)(14\Theta_1^2+16\Theta_1\Theta_2+5\Theta_2^2+16\Theta_1+10\Theta_2+5),
\label{PFdet2}
\end{eqnarray}
where $\Theta_{\ell}=z_{\ell}\frac{\partial}{\partial z_{\ell}}$. These operators can be also derived by the method of \cite{Hosono:1994ax,Hosono:2011np,Hosono:2012hc}. Following them, first we can obtain the Picard-Fuchs operators $\widehat{\cal D}_1=\Theta_1^6-z_1(\Theta_1+\Theta_2+1)^6$ and $\widehat{\cal D}_2=\Theta_2^6-z_2(\Theta_1+\Theta_2+1)^6$ from the charge assignment in Table \ref{PAX1}. Then the above operators (\ref{PFdet1}) and (\ref{PFdet2}) are derived from the irreducible factorizations as
\begin{eqnarray}
&&
280\widehat{\cal D}_1=
(\Theta_1+\Theta_2)(14\Theta_1^2+16\Theta_1\Theta_2+5\Theta_2^2){\cal D}_1+(42\Theta_1^3+54\Theta_1^2\Theta_2+27\Theta_1\Theta_2^2+5\Theta_2^3){\cal D}_2,\nonumber\\
&&\\
&&
280\widehat{\cal D}_2=
(\Theta_1+\Theta_2)(5\Theta_1^2+16\Theta_1\Theta_2+14\Theta_2^2){\cal D}_1-(5\Theta_1^3+27\Theta_1^2\Theta_2+54\Theta_1\Theta_2^2+42\Theta_2^3){\cal D}_2.\nonumber\\
&&
\end{eqnarray}

Using the above results (\ref{detp0}) -- (\ref{detG12}), we have checked that the conjectural forms (\ref{PF4c}) are annihilated by the Picard-Fuchs operators (\ref{PFdet1}) and (\ref{PFdet2}) up to $z_1^4$ and $z_2^4$.

\subsection{Complete intersection in Grassmannian: $X_{1^8}\subset G(2,8)$}

As an example with non-abelian GLSM description, we consider the Grassmannian Calabi-Yau fourfold $X_{1^8}\subset G(2,8)$ defined by the complete intersection of eight hyperplanes with degree one in the Grassmannian $G(2,8)$. First we compute the classical topological invariants of this manifold. We denote the class of a hyperplane section as $\sigma_1=c_1(Q)$, where $Q$ is the universal quotient bundle of $G(2,8)$. Then the classical quadruple intersection number is calculated as $\kappa=\int_{X_{1^8}}\sigma_1^4=132$, and we also see that $\int_{X_{1^8}}c_3(X_{1^8})\wedge \sigma_1=-336$ (for details, see Appendix A).\footnote{The mirror construction of Calabi-Yau complete intersections in $G(k,n)$ was studied in \cite{Batyrev:1998kx} by making use of a flat deformation of Grassmannian $G(k,n)$ to a Gorenstein toric Fano variety \cite{Sturmfels:1996}.}

\begin{table}[t]
\begin{center}
\begin{tabular}{c|c|c}
\hline
Field & $U(2)$  & $U(1)_V$ \\ \hline
$\Phi^i$ & $\mathbf{2}_{+1}$ & $2\mathfrak{q}$ \\
$P_a$ & $\mathbf{1}_{-2}$ & $2-4\mathfrak{q}$ \\ \hline
\end{tabular}
\caption{Matter content of the $U(2)$ GLSM describing the complete intersection Calabi-Yau in Grassmannian $X_{1^8}\subset G(2,8)$. Here $i,a=1,\ldots,8$. The subscript denotes the charge under the central $U(1)\subset U(2)$. R-charges are assigned so that the total R-charge of superpotential is $2$, and by the positivity constraint, $0<\mathfrak{q}<\frac12$.}
\label{cicy}
\end{center}
\end{table}

The GLSM which describes $X_{1^8}\subset G(2,8)$ has $U(2)$ gauge group and matter multiplets given in Table \ref{cicy} (see \cite{Hori:2006dk}). These chiral multiplets are coupled through a superpotential $W=\overset{8}{\underset{{a,i,j=1}}{\sum}}
A^a_{ij}P_a(\Phi_1^i\Phi_2^j-\Phi_2^i\Phi_1^j)$, where $A^a_{ij}$ are eight antisymmetric $8\times 8$ matrices.

Using the formulas (\ref{GLSMp}) -- (\ref{GLSMmatt}), the partition function of this model is given by
\begin{eqnarray}
Z_{\mbox{\scriptsize GLSM}}&=&
\frac12\sum_{m_1,m_2\in{\IZ}}e^{-i\theta (m_1+m_2)}\int_{-\infty}^{\infty}\frac{d\sigma_1}{2\pi}\frac{d\sigma_2}{2\pi}e^{-4\pi ir(\sigma_1+\sigma_2)}\bigg[\frac{(m_1-m_2)^2}{4}+(\sigma_1-\sigma_2)^2\bigg]\nonumber\\
&&\hspace{-3.8em}
\times\bigg[\frac{\Gamma(\mathfrak{q}-i\sigma_1-\frac{1}{2}m_1)}{\Gamma(1-\mathfrak{q}+i\sigma_1-\frac{1}{2}m_1)}\frac{\Gamma(\mathfrak{q}-i\sigma_2-\frac{1}{2}m_2)}{\Gamma(1-\mathfrak{q}+i\sigma_2-\frac{1}{2}m_2)}\frac{\Gamma\big(1-2\mathfrak{q}+i(\sigma_1+\sigma_2)+\frac12(m_1+m_2)\big)}{\Gamma\big(2\mathfrak{q}-i(\sigma_1+\sigma_2)+\frac12(m_1+m_2)\big)}\bigg]^8.\nonumber\\
&&
\end{eqnarray}
Here we consider the Grassmann phase $r \gg 0$ which corresponds to the non-linear sigma model on $X_{1^8}\subset G(2,8)$. In this phase, the partition function can be evaluated as
\begin{align}
Z_{\mbox{\scriptsize GLSM}}=&
-\frac12(z\overline{z})^{2\mathfrak{q}}\oint\frac{d\epsilon_1}{2\pi i}\frac{d\epsilon_2}{2\pi i}(z\overline{z})^{-\epsilon_1-\epsilon_2}
\frac{\pi^8\sin^8\pi(\epsilon_1+\epsilon_2)}{\sin^8(\pi \epsilon_1)\sin^8(\pi \epsilon_2)}\nonumber\\
&
\times\bigg|\sum_{k_1,k_2=0}^{\infty}z^{k_1+k_2}\big[(k_1-k_2)-(\epsilon_1-\epsilon_2)\big]\frac{\Gamma(1+(k_1+k_2)-(\epsilon_1+\epsilon_2))^8}{\Gamma(1+k_1-\epsilon_1)^8\Gamma(1+k_2-\epsilon_2)^8}\bigg|^2,
\end{align}
where $z=e^{-2\pi r+i\theta}$, and the complex conjugation does not act on $\epsilon_{1,2}$. 

The coefficient of $\log^4\overline{z}$ is given by
\begin{align}
\hspace{-4em}
\frac{132}{4!}(z\overline{z})^{2\mathfrak{q}}T^0(z)\overline{T^0(z)},
\end{align}
where
\begin{align}
T^0(z)&=\sum_{k_1,k_2=0}^{\infty}z^{k_1+k_2}\frac{((k_1+k_2)!)^8}{(k_1!k_2!)^8}\big[1-4(k_1-k_2)\big(\Psi(1+k_1)-\Psi(1+k_2)\big)\big]\nonumber\\
&=1-6z+234z^2-13164z^3+936810z^4-76041756z^5+\cdots,
\label{GLSMGrcoeff}
\end{align}
We see that the above series expansion of $T^0(-z)$ agrees with the fundamental period
\begin{equation}
\widehat{T}^0(z)=
\sum_{\ell_0,\ell_1,\ldots,\ell_5=0}^{\infty}z^{\ell_0}\binom {\ell_0}{\ell_1}\binom {\ell_2}{\ell_1}\binom {\ell_0}{\ell_2}
\binom {\ell_3}{\ell_2}\binom {\ell_0}{\ell_3}\binom {\ell_4}{\ell_3}\binom {\ell_0}{\ell_4}\binom {\ell_5}{\ell_4}\binom {\ell_0}{\ell_5}^2
\label{Grass4Peri}
\end{equation}
of the mirror manifold of $X_{1^8}\subset G(2,8)$ which can be obtained by means of a prescription of \cite{Batyrev:1998kx} (see also \cite{Batyrev:1998mit}). In Appendix A, we will revisit this issue and refer to a generalization. As noted in \cite{Jockers:2012dk}, it is interesting to prove this type of coincidence from the viewpoint of combinatorics.

After normalizing the partition function $Z_{\mbox{\scriptsize GLSM}}$ by $(z\overline{z})^\mathfrak{q}(2\pi i)^4T^0\overline{T^0}$, we can determine the flat coordinate by picking up the coefficient of $\log^3\overline{z}$. The result is given by
\begin{eqnarray}
2\pi it&=&\log z+2\pi it_{(0)}
-\frac{4}{T^0}\sum_{k_1,k_2=0}^{\infty}\frac{((k_1+k_2)!)^8}{(k_1!)^8(k_2!)^8}
z^{k_1+k_2}
\Big[(k_1-k_2)\Psi^{(1)}(1+k_1)\hspace{5.5em}\nonumber\\
&&
-2\big[\Psi(1+k_1+k_2)-\Psi(1+k_1)\big]
\big[1-4(k_1-k_2)\big(\Psi(1+k_1)-\Psi(1+k_2)\big)\big]\Big],
\end{eqnarray}
where $\Psi^{(1)}(x)=\frac{d}{dx}\Psi(x)$. By taking $t_{(0)}=\frac12$, we find that the generating function (\ref{HolGij}) associated with $H_1\equiv\sigma_1^2$ is given by
\begin{equation}
F_1(t)\equiv G(t)=\frac{132}{2}t^2+\frac{1}{(2\pi i)^2}\sum_{d=1}^{\infty}n_d{\rm Li}_2(q^d),
\label{Grgene}
\end{equation}
where the Gromov-Witten invariants $n_d$ are listed in Table \ref{cicygrGW}.
\begin{table}[t]
\begin{center}
\begin{tabular}{c|r|r}
\hline
$d$ & $n_d$ & $m_d$\\ \hline
1 & 1680 & 672 \\
2 & 50904  & 3360 \\
3 & 2003568  & 156352 \\
4 & 108147648  & 7928256 \\
5 & 6684193824  & 482638464 \\
6 & 456302632296  & 32296103952 \\ 
7 & 33294956299248  & 2327924504640 \\ 
8 & 2553533188012800  & 176807420598144 \\  \hline
\end{tabular}
\caption{Gromov-Witten invariants for $X_{1^8}\subset G(2,8)$.}
\label{cicygrGW}
\end{center}
\end{table}

In this example $\dim H^{2,2}_{\mbox{\scriptsize prim}}(X_{1^8})=2$, and we can take an element orthogonal to $H_1$ as $H_2\equiv 22\sigma_2-15\sigma_1^2$, where $\sigma_2=c_2(Q)$ (see Appendix A).\footnote{In the previous version of this paper, we did not find the explicit form of $H_2$. We would grateful to the referee for pointing out this missing.} The intersection matrix in this basis $\{H_1,H_2\}$ is given by
\begin{equation}
\eta_{mn}=\rm{diag}(132,\ 308).
\end{equation}
Then we can also obtain a generating function of Gromov-Witten invariants (\ref{HolFnpr}) associated with the element $H_2$:
\begin{equation}
F_2(t)=\frac{1}{(2\pi i)^2}\sum_{d=1}^{\infty}m_d{\rm Li}_2(q^d).
\label{Grass_resG}
\end{equation}
Here the Gromov-Witten invariants $m_d$ are shown in Table \ref{cicygrGW}. Combining these results, we have also checked our conjecture (\ref{Kahl4}) holds in this example. Note that the Gromov-Witten invariants in Table \ref{cicygrGW} are the predictions based on the relation (\ref{JKLMR2}) and our conjecture (\ref{Kahl4}).

Let us reconsider the problem from the viewpoint of mirror symmetry. We can check that the fundamental period (\ref{Grass4Peri}) is a kernel of the following Picard-Fuchs operator
\begin{align}
{\cal D}= \ &
121(\Theta-1)\Theta^5-22z(438\Theta^5+2094\Theta^4+1710\Theta^3+950\Theta^2+275\Theta+33)\Theta\nonumber\\
&
-z^2(839313\Theta^6+2471661\Theta^5+4037556\Theta^4+4497304\Theta^3+3093948\Theta^2+1158740\Theta \nonumber \\
&+180048)-2z^3(5746754\Theta^6+26470666\Theta^5+51184224\Theta^4+50480470\Theta^3 \nonumber \\
&+26295335\Theta^2+6684843\Theta+604098)-4z^4(4081884\Theta^6+14894484\Theta^5 \nonumber \\
&+18825903\Theta^4+7472030\Theta^3-3698839\Theta^2-4099839\Theta-993618)+56z^5(29592\Theta^6 \nonumber \\
&+255960\Theta^5+806448\Theta^4+1272787\Theta^3+1088403\Theta^2+483431\Theta+87609) \nonumber \\
&+1568z^6(4\Theta+5)(2\Theta+3)(4\Theta+3)(\Theta+1)^3.
\label{PF_Grass8}
\end{align}
We have checked that (\ref{PF4c}) are precisely the kernels of the Picard-Fuchs operator (\ref{PF_Grass8}) up to $z^8$.

\section{Local toric Calabi-Yau varieties}

In this section, we consider $d$ dimensional local toric Calabi-Yau varieties with K\"ahler parameters $r_{\ell}$. Here $\ell =1, \ldots, n-d$. Each of these varieties has an ${\cal N}=(2,2)$ abelian GLSM description and can be constructed by the symplectic quotient
\begin{equation}
X=\Big\{(\phi_1,\ldots,\phi_n)\in {\IC}^n \Big| \sum_{i=1}^n Q_i^{\ell}|\phi_i|^2=r_{\ell},~\sum_{i=1}^n Q_i^{\ell}=0\Big\}/U(1)^{n-d},
\end{equation}
where $Q_i^{\ell} \in{\IZ}$ are $n-d$ charge vectors with $n$ components. The $U(1)^{n-d}$ gauge group acts on the complex scalar fields $\phi_i$ as $\phi_i \to e^{i \sum_{\ell}\epsilon_{\ell}Q_i^{\ell}}\phi_i$.

Let us consider whether the relation (\ref{JKLMR2}) proposed in \cite{Jockers:2012dk} can be applied to the local toric Calabi-Yau cases. In fact, by taking the geometric engineering limit \cite{Katz:1996fh} for the local Hirzebruch surface ${\cal O}(-2,-2)\to {\IP}^1\times{\IP}^1$, the Seiberg-Witten K\"ahler potential for ${\cal N}=2$ pure $SU(2)$ SYM has been obtained from the GLSM calculation in \cite{Park:2012nn} (see also Section 5.1.2). Due to the non-compactness of local toric varieties, as suggested in \cite{Park:2012nn}, we need to take all the R-charges of the chiral superfields in the corresponding GLSM to be zero. Although the partition function $Z_{\mbox{\scriptsize GLSM}}$ on $S^2$ diverges under this ``non-compact limit'', the authors of \cite{Park:2012nn} claimed that the correct Seiberg-Witten K\"ahler potential can be obtained from the regular part of $Z_{\mbox{\scriptsize GLSM}}$. As a generalization of their result, here we claim that the K\"ahler potential $K$ on the K\"ahler moduli space of local toric Calabi-Yau varieties can be obtained by
\begin{equation}
e^{-K}=\oint\frac{d\mathfrak{q}_1}{2\pi i\mathfrak{q}_1}\cdots\oint\frac{d\mathfrak{q}_m}{2\pi i\mathfrak{q}_m}Z_{\mbox{\scriptsize GLSM}},
\label{toricclaim}
\end{equation}
up to the degrees of freedom of K\"ahler transformation. $\{\mathfrak{q}_i\}_{i=1}^m$ are the R-charges of $m$ chiral superfields corresponding to the non-compact directions. In the remaining part of this section, we will check our claim in several examples of threefolds and fourfolds. During the computation, we treat the R-charges of the chiral superfields related to the non-compact directions as regulators of the divergence.

\subsection{Threefolds}

Here we consider local toric Calabi-Yau threefolds. By performing the exact calculation of GLSM partition functions and using the explicit form of the K\"ahler potential (\ref{3dKahl}), we can study their topological nature and also confirm the consistency of our claim (\ref{toricclaim}).

\subsubsection{Resolved conifold: ${\cal O}(-1)\oplus{\cal O}(-1)\to {\IP}^1$}

The GLSM corresponding to the resolved conifold ${\cal O}(-1)\oplus{\cal O}(-1)\to {\IP}^1$ can be characterized by the charge vector $Q=(-1,-1,1,1)$. 
Assigning the same R-charge $2\mathfrak{q}$ to two chiral superfields corresponding to two non-compact directions, the GLSM partition function (\ref{GLSMp}) is evaluated as
\begin{eqnarray}
Z_{\mbox{\scriptsize GLSM}}&=&
\sum_{m\in{\IZ}}e^{-i\theta m}\int_{-\infty}^{\infty}\frac{d\sigma}{2\pi}e^{-4\pi ir\sigma}
\frac{\Gamma(-i\sigma-\frac{1}{2}m)^2}{\Gamma(1+i\sigma-\frac{1}{2}m)^2}\frac{\Gamma(\mathfrak{q}+i\sigma+\frac{1}{2}m)^2}{\Gamma(1-\mathfrak{q}-i\sigma+\frac{1}{2}m)^2}\nonumber\\
&=&
\oint\frac{d\epsilon}{2\pi i}(z\overline{z})^{-\epsilon}
\frac{\sin^2\pi(\mathfrak{q}-\epsilon)}{\sin^2(\pi \epsilon)}
\bigg|\sum_{k=0}^{\infty}z^k\frac{\Gamma(\mathfrak{q}+k-\epsilon)^2}{\Gamma(1+k-\epsilon)^2}\bigg|^2,
\end{eqnarray}
where $z=e^{-2\pi r+i\theta}$ and the complex conjugation does not act on $\epsilon$. 

First we normalize the partition function as
\begin{equation}
\widetilde{Z}_{\mbox{\scriptsize GLSM}}=\frac{g(z)\overline{g(z)}}{f(z)\overline{f(z)}}Z_{\mbox{\scriptsize GLSM}}.
\end{equation}
Here the normalization factors are
\begin{eqnarray}
&&
f(z)=
\sum_{k=0}^{\infty}\frac{\Gamma(\mathfrak{q}+k)^2}{\Gamma(1+k)^2}z^k,\\
&&
g(z)=
\widetilde{\Gamma}(\mathfrak{q}, h(z))^2
\bigg[1+\mathfrak{q}^2\sum_{k=1}^{\infty}\frac{z^k}{\Gamma(1+k)^2}\prod_{j=1}^{k-1}(\mathfrak{q}+j)^2\bigg],
\end{eqnarray}
where
\begin{align}
\widetilde{\Gamma}(\mathfrak{q},h(z))=\sum_{n=0}^{\infty}\frac{\Gamma^{(n)}(1)}{n!}\bigg|_{\gamma=h(z)}\mathfrak{q}^{n-1},
\end{align}
and $\gamma$ is the Euler constant. Note that $\widetilde{\Gamma}(\mathfrak{q}, \gamma)=\Gamma(\mathfrak{q})$, and thus $g(z)\big|_{h(z)=\gamma}=f(z)$. Under the non-compact limit $\mathfrak{q}\to 0^+$ \cite{Park:2012nn}, this normalization only replaces $\gamma$ with a holomorphic function $h(z)$. 
This prescription is necessary to produce the classical term of the K\"ahler potential. Then the behavior of the partition function under the non-compact limit is given by
\begin{align}
\widetilde{Z}_{\mbox{\scriptsize GLSM}}=& \ 
2\mathfrak{q}^{-3}-\big(4h(z)+4\overline{h(z)}+\log z\overline{z}\big)\mathfrak{q}^{-2}\nonumber\\
&
+2\big(h(z)+\overline{h(z)}\big)\big(2h(z)+2\overline{h(z)}+\log z\overline{z}\big)\mathfrak{q}^{-1}
+\widetilde{Z}_0+{\cal O}(\mathfrak{q}),
\end{align}
where
\begin{align}
\widetilde{Z}_0
=&-\frac83\big(h(z)+\overline{h(z)}\big)^3-2\big(h(z)+\overline{h(z)}\big)^2\log z\overline{z}+\frac43\zeta(3)\nonumber\\
&
-\big({\rm Li}_2(z)+{\rm Li}_2(\overline{z})\big)\log z\overline{z}+2\big({\rm Li}_3(z)+{\rm Li}_3(\overline{z})\big).
\label{resZq}
\end{align}
By taking $h(z)=-\frac14 \log z$, we find that (\ref{resZq}) gives the K\"ahler potential of the form (\ref{3dKahl}) with the ``natural classical triple intersection number'' $\kappa=\frac12$ \cite{Forbes:2005xt}. This result is consistent with our statement (\ref{toricclaim}). In a similar way to the case of the compact Calabi-Yau in \cite{Jockers:2012dk}, we can also read off the flat coordinate $t$ and the Gromov-Witten invariants $n_d$ as
\begin{equation}
2\pi i t=\log z,\ \ n_1=1,~n_{d\ge 2}=0.
\end{equation}

\subsubsection{Local Hirzebruch surface: ${\cal O}(-2,-2)\to {\IP}^1\times{\IP}^1$}

Let us consider the local Hirzebruch surface ${\cal O}(-2,-2)\to {\IP}^1\times{\IP}^1$ defined by two charge vectors $Q^1=(-2,1,1,0,0)$ and $Q^2=(-2,0,0,1,1)$. The exact partition function of the corresponding GLSM with an R-charge $\mathfrak{q}$ is given by
\begin{eqnarray}
Z_{\mbox{\scriptsize GLSM}}&=&
\sum_{m_1,m_2\in{\IZ}}e^{-i(\theta_1m_1+\theta_2m_2)}\int_{-\infty}^{\infty}\frac{d\sigma_1}{2\pi}\frac{d\sigma_2}{2\pi}e^{-4\pi i(r_1\sigma_1+r_2\sigma_2)}\nonumber\\
&&
\times\frac{\Gamma(-i\sigma_1-\frac{1}{2}m_1)^2}{\Gamma(1+i\sigma_1-\frac{1}{2}m_1)^2}\frac{\Gamma(-i\sigma_2-\frac{1}{2}m_2)^2}{\Gamma(1+i\sigma_2-\frac{1}{2}m_2)^2}\frac{\Gamma(\mathfrak{q}+2i(\sigma_1+\sigma_2)+(m_1+m_2))}{\Gamma(1-\mathfrak{q}-2i(\sigma_1+\sigma_2)+(m_1+m_2))}\nonumber\\
&=&
\oint\frac{d\epsilon_1}{2\pi i}\frac{d\epsilon_2}{2\pi i}(z_1\overline{z}_1)^{-\epsilon_1}(z_2\overline{z}_2)^{-\epsilon_2}
\frac{\pi^3\sin\pi\big(\mathfrak{q}-2(\epsilon_1+\epsilon_2)\big)}{\sin^2(\pi \epsilon_1)\sin^2(\pi \epsilon_2)}\nonumber\\
&&\hspace{8em}
\times\bigg|\sum_{k_1,k_2=0}^{\infty}z_1^{k_1}z_2^{k_2}\frac{\Gamma(\mathfrak{q}+2(k_1+k_2)-2(\epsilon_1+\epsilon_2))}{\Gamma(1+k_1-\epsilon_1)^2\Gamma(1+k_2-\epsilon_2)^2}\bigg|^2,
\end{eqnarray}
where $z_{\ell}=e^{-2\pi r_{\ell}+i\theta_{\ell}}$ and the complex conjugation does not act on $\epsilon_{1,2}$. As in the case of the resolved conifold, we take the non-compact limit $\mathfrak{q}\to 0^+$ of a normalized partition function $\widetilde{Z}_{\mbox{\scriptsize GLSM}}=\frac{g(z_1,z_2)\overline{g(z_1,z_2)}}{f(z_1,z_2)\overline{f(z_1,z_2)}}Z_{\mbox{\scriptsize GLSM}}$, where
\begin{eqnarray}
f(z_1,z_2)&=&
\sum_{k_1,k_2=0}^{\infty}\frac{\Gamma(\mathfrak{q}+2(k_1+k_2))}{\Gamma(1+k_1)^2\Gamma(1+k_2)^2}z_1^{k_1}z_2^{k_2},\hspace{15em}\\
g(z_1,z_2)&=&
\widetilde{\Gamma}(\mathfrak{q}, h(z))
\bigg[1+\mathfrak{q}\mathop{\sum_{k_1,k_2=0}^{\infty}}\limits_{(k_1,k_2)\ne (0,0)}\frac{z_1^{k_1}z_2^{k_2}}{\Gamma(1+k_1)^2\Gamma(1+k_2)^2}\prod_{j=1}^{2k_1+2k_2-1}(\mathfrak{q}+j)\bigg],
\end{eqnarray}
and $h(z) \equiv h(z_1,z_2)$ is a holomorphic function of $z_{1,2}$. Then we obtain the Laurent expansion
\begin{eqnarray}
\widetilde{Z}_{\mbox{\scriptsize GLSM}}&=&
8\mathfrak{q}^{-3}-2\big(4h(z)+4\overline{h(z)}+\log z_1z_2+\log\overline{z}_1\overline{z}_2\big)\mathfrak{q}^{-2}\hspace{10.5em}\nonumber\\
&&\hspace{-1em}
+\big(2h(z)+2\overline{h(z)}+\log z_1\overline{z}_1\big)\big(2h(z)+2\overline{h(z)}+\log z_2\overline{z}_2\big)\mathfrak{q}^{-1}+\widetilde{Z}_0+{\cal O}(\mathfrak{q}),
\end{eqnarray}
where $\widetilde{Z}_0$ is given by
\begin{eqnarray}
&&
\widetilde{Z}_0=
-\frac43\big(h(z)+\overline{h(z)}\big)^3-\big(h(z)+\overline{h(z)}\big)^2\log z_1\overline{z}_1z_2\overline{z}_2-\big(h(z)+\overline{h(z)}\big)\log z_1\overline{z}_1\log z_2\overline{z}_2\nonumber\\
&&\hspace{1em}
+\big(\Delta_{00}(z)+\overline{\Delta_{00}(z)}\big)\log z_1\overline{z}_1\log z_2\overline{z}_2+2\Delta_{00}(z)\overline{\Delta_{00}(z)}\log z_1\overline{z}_1z_2\overline{z}_2-\frac{16}{3}\zeta(3)\nonumber\\
&&\hspace{1em}
+\big(\Delta_{01}(z)+\overline{\Delta_{01}(z)}\big)\log z_1\overline{z}_1+\big(\Delta_{10}(z)+\overline{\Delta_{10}(z)}\big)\log z_2\overline{z}_2-\frac23\pi^2\big(\Delta_{00}(z)+\overline{\Delta_{00}(z)}\big)\nonumber\\
&&\hspace{1em}
+2\Delta_{00}(z)\big(\overline{\Delta_{10}(z)}+\overline{\Delta_{01}(z)}\big)+2\overline{\Delta_{00}(z)}\big(\Delta_{10}(z)+\Delta_{01}(z)\big)+\Delta_{11}(z)+\overline{\Delta_{11}(z)}.
\label{hirzZq}
\end{eqnarray}
In the above expression, we have defined $\Delta_{00}(z)\equiv\Delta_{00}(z_1,z_2)$, $\Delta_{10}(z)\equiv\Delta_{10}(z_1,z_2)$, $\Delta_{01}(z)\equiv\Delta_{01}(z_1,z_2)=\Delta_{10}(z_2,z_1)$, and $\Delta_{11}(z)\equiv\Delta_{11}(z_1,z_2)$ as
\begin{eqnarray}
\Delta_{00}(z)&=&\mathop{\sum_{k_1,k_2=0}^{\infty}}\limits_{(k_1,k_2)\ne (0,0)}\frac{(2k_1+2k_2-1)!}{(k_1!)^2(k_2!)^2}z_1^{k_1}z_2^{k_2},\\
\Delta_{10}(z)&=&2\mathop{\sum_{k_1,k_2=0}^{\infty}}\limits_{(k_1,k_2)\ne (0,0)}\frac{(2k_1+2k_2-1)!}{(k_1!)^2(k_2!)^2}z_1^{k_1}z_2^{k_2}\big[\Psi(2k_1+2k_2)-\Psi(1+k_1)\big],\\
\Delta_{11}(z)&=&4\mathop{\sum_{k_1,k_2=0}^{\infty}}\limits_{(k_1,k_2)\ne (0,0)}\frac{(2k_1+2k_2-1)!}{(k_1!)^2(k_2!)^2}z_1^{k_1}z_2^{k_2}\Big[\Psi^{(1)}(2k_1+2k_2)\nonumber\\
&&\hspace{2em}
+\big[\Psi(2k_1+2k_2)-\Psi(1+k_1)\big]\big[\Psi(2k_1+2k_2)-\Psi(1+k_2)\big]\Big],
\end{eqnarray}
where $\Psi(x)=\frac{d}{dx}\log\Gamma(x)$ and $\Psi^{(1)}(x)=\frac{d}{dx}\Psi(x)$. By choosing $h(z)=-\frac14 \log (z_1z_2)$ and comparing (\ref{hirzZq}) with (\ref{3dKahl}), we obtain the classical triple intersection numbers $\kappa_{111}=\kappa_{222}=\frac14$ and $\kappa_{112}=\kappa_{122}=-\frac14$ computed in \cite{Forbes:2005xt,Haghighat:2008gw}. The flat coordinates are given by
\begin{equation}
2\pi i t^1=\log z_1-2\Delta_{00}(z),\ \
2\pi i t^2=\log z_2-2\Delta_{00}(z),
\end{equation}
and we finally obtain
\begin{eqnarray}
\frac{i}{(2\pi i)^3}\widetilde{Z}_0&=&
-\frac{i}{6}\sum_{\ell,m,n=1,2}\kappa_{\ell mn}(t^{\ell}-{\overline t}^{\ell})(t^m-{\overline t}^m)(t^n-{\overline t}^n)+\frac{2}{3\pi^3}\zeta(3)\hspace{9.7em}\nonumber\\
&&\hspace{-3em}
-\frac{i}{(2\pi i)^2}\big(\Delta_{00}(z)^2-\Delta_{01}(z)+\overline{\Delta_{00}(z)}^2-\overline{\Delta_{01}(z)}\big)(t^1-\overline{t}^1)\nonumber\\
&&\hspace{-3em}
-\frac{i}{(2\pi i)^2}\big(\Delta_{00}(z)^2-\Delta_{10}(z)+\overline{\Delta_{00}(z)}^2-\overline{\Delta_{10}(z)}\big)(t^2-\overline{t}^2)\nonumber\\
&&\hspace{-3em}
+\frac{2i}{(2\pi i)^3}\Big(\frac43\Delta_{00}(z)^3-\Delta_{00}(z)\big(\Delta_{10}(z)+\Delta_{01}(z)\big)+\Delta_{11}(z)-\frac{\pi^2}{3}\Delta_{00}(z)+c.c.\Big).
\end{eqnarray}
From this result, we can extract the Gromov-Witten invariants $n_{d_1,d_2}=n_{d_2,d_1}$ as
\begin{equation}
n_{1,0}=-2,~n_{2,0}=0,~n_{1,1}=-4,~n_{2,1}=-6,~n_{3,1}=-8,~n_{2,2}=-32,~n_{3,2}=-110,\ldots\ .
\end{equation}
These are in agreement with the computation in \cite{Chiang:1999tz}.

\subsection{Fourfolds}

Next, we study local toric Calabi-Yau fourfolds. Just like the case of threefolds, using (\ref{Kahl4}) and (\ref{toricclaim}) we compute the Gromov-Witten invariants for three local examples discussed in \cite{Klemm:2007in}. This also corresponds to the nontrivial check for our statements.

\subsubsection{Local ${\IP}^2$: ${\cal O}(-1)\oplus{\cal O}(-2)\to {\IP}^2$}

Toric charge of the local Calabi-Yau ${\cal O}(-1)\oplus{\cal O}(-2)\to {\IP}^2$ is given by $Q=(-1,-2,1,1,1)$, and we denote the K\"ahler form defined on the base ${\IP}^2$ by $J$. Using this data, we can construct GLSM and the partition function is evaluated as
\begin{eqnarray}
Z_{\mbox{\scriptsize GLSM}}&=&
\sum_{m\in{\IZ}}e^{-i\theta m}\int_{-\infty}^{\infty}\frac{d\sigma}{2\pi}e^{-4\pi ir\sigma}\nonumber\\
&&
\times\frac{\Gamma(-i\sigma-\frac{1}{2}m)^3}{\Gamma(1+i\sigma-\frac{1}{2}m)^3}\frac{\Gamma(\mathfrak{q}_1+i\sigma+\frac{1}{2}m)}{\Gamma(1-\mathfrak{q}_1-i\sigma+\frac{1}{2}m)}\frac{\Gamma(\mathfrak{q}_2+2i\sigma+m)}{\Gamma(1-\mathfrak{q}_2-2i\sigma+m)}\nonumber\\
&=&
\oint\frac{d\epsilon}{2\pi i}(z\overline{z})^{-\epsilon}
\frac{\pi\sin\pi(\mathfrak{q}_1-\epsilon)\sin\pi(\mathfrak{q}_2-2\epsilon)}{\sin^3(\pi \epsilon)}\nonumber\\
&&\hspace{8em}
\times\bigg|\sum_{k=0}^{\infty}(-z)^{k}\frac{\Gamma(\mathfrak{q}_1+k-\epsilon)\Gamma(\mathfrak{q}_2+2k-2\epsilon)}{\Gamma(1+k-\epsilon)^3}\bigg|^2.
\end{eqnarray}
Here $z=e^{-2\pi r+i\theta}$ and the complex conjugation does not act on $\epsilon_{1,2}$. In a similar way to the above threefold examples, we consider asymptotic behavior of a normalized partition function $\widetilde{Z}_{\mbox{\scriptsize GLSM}}=\frac{g(z)\overline{g(z)}}{f(z)\overline{f(z)}}Z_{\mbox{\scriptsize GLSM}}$ under the non-compact limit $\mathfrak{q}_{1,2}\to 0^+$. Here $f(z)$ and $g(z)$ are defined by
\begin{eqnarray}
f(z)&=&
\sum_{k=0}^{\infty}\frac{\Gamma(\mathfrak{q}_1+k)\Gamma(\mathfrak{q}_2+2k)}{\Gamma(1+k)^3}(-z)^{k},\hspace{21.6em}\\
g(z)&=&
\widetilde{\Gamma}(\mathfrak{q}_1, h(z))\widetilde{\Gamma}(\mathfrak{q}_2, h(z))
\bigg[1+\mathfrak{q}_1\mathfrak{q}_2\sum_{k=1}^{\infty}\frac{(-z)^{k}}{\Gamma(1+k)^3}\prod_{j_1=1}^{k-1}(\mathfrak{q}_1+j_1)\cdot\prod_{j_2=1}^{2k-1}(\mathfrak{q}_2+j_2)\bigg].
\end{eqnarray}
Performing the double series expansion, we obtain
\begin{eqnarray}
&&
\widetilde{Z}_{\mbox{\scriptsize GLSM}}=\frac{(2\mathfrak{q}_1+\mathfrak{q}_2)^2}{\mathfrak{q}_1^{3}\mathfrak{q}_2^{3}}-\big(h(z)+\overline{h(z)}\big)\frac{4\mathfrak{q}_1^3+\mathfrak{q}_2^3}{\mathfrak{q}_1^{3}\mathfrak{q}_2^{3}}
-\big(3h(z)+3\overline{h(z)}+\log z\overline{z}\big)\frac{2\mathfrak{q}_1+\mathfrak{q}_2}{\mathfrak{q}_1^{2}\mathfrak{q}_2^{2}}\hspace{2.2em}\nonumber\\
&&\hspace{1.5em}
+\big(h(z)+\overline{h(z)}\big)\big(2h(z)+2\overline{h(z)}+\log z\overline{z}\big)\mathfrak{q}_1^{-2}+\frac12\big(3h(z)+3\overline{h(z)}+\log z\overline{z}\big)^2\mathfrak{q}_1^{-1}\mathfrak{q}_2^{-1}\nonumber\\
&&\hspace{1.5em}
+\big(h(z)+\overline{h(z)}\big)\big(5h(z)+5\overline{h(z)}+2\log z\overline{z}\big)\mathfrak{q}_2^{-2}+\widetilde{Z}_{10}\mathfrak{q}_1^{-1}+\widetilde{Z}_{01}\mathfrak{q}_2^{-1}+\widetilde{Z}_{00}+\cdots,
\end{eqnarray}
where
\begin{align}
\widetilde{Z}_{10}-\widetilde{Z}_{01} \ \ &=\big(h(z)+\overline{h(z)}\big)^3+\frac12\big(h(z)+\overline{h(z)}\big)^2\log z\overline{z}-2\zeta(3),\\
\widetilde{Z}_{10}-4\widetilde{Z}_{01}&=\frac12\big(h(z)+\overline{h(z)}\big)\big(3h(z)+3\overline{h(z)}+\log z\overline{z}\big)\big(7h(z)+7\overline{h(z)}+3\log z\overline{z}\big),
\end{align}
and
\begin{eqnarray}
\widetilde{Z}_{00}&=&
\frac43\big(h(z)+\overline{h(z)}\big)^4+\frac32\big(h(z)+\overline{h(z)}\big)^3\log z\overline{z}+\big(h(z)+\overline{h(z)}\big)^2(\log z\overline{z})^2\nonumber\\
&&
+\frac{10}{3}\zeta(3)\big(h(z)+\overline{h(z)}\big)+\frac12\big(\Delta_0(z)+\overline{\Delta_0(z)}\big)(\log z\overline{z})^2-\big(\Delta_1(z)+\overline{\Delta_1(z)}\big)\log z\overline{z}\nonumber\\
&&
+2\Delta_0(z)\overline{\Delta_0(z)}+\Delta_2(z)+\overline{\Delta_2(z)}-\frac{1}{6}\pi^2\big(\Delta_0(z)+\overline{\Delta_0(z)}\big).
\label{localA1Zq}
\end{eqnarray}
In the above expression, we have defined
\begin{align}
\Delta_0(z)&=\sum_{k=1}^{\infty}\frac{(-1)^k}{2k^2}\binom {2k}{k}z^k,\\
\Delta_1(z)&=\sum_{k=1}^{\infty}\frac{(-1)^k}{2k^3}\binom {2k}{k}z^k
-2\sum_{k=1}^{\infty}\frac{(-1)^k}{2k^2}\binom {2k}{k}z^k\big[\Psi(2k)-\Psi(1+k)\big],\\
\Delta_2(z)&=\frac12\sum_{k=1}^{\infty}\frac{(-1)^k}{2k^2}\binom {2k}{k}z^k
\Big[\big[\Psi(k)+2\Psi(2k)-3\Psi(1+k)\big]^2\nonumber\\
&\hspace{10em}
+\Psi^{(1)}(k)+4\Psi^{(1)}(2k)-3\Psi^{(1)}(1+k)\Big].
\end{align}
Let us choose $h(z)=-\frac12 \log z$. By comparing (\ref{localA1Zq}) with our conjecture (\ref{Kahl4}) for the K\"ahler potential, we can obtain the ``classical quadruple intersection number'' $\kappa=\frac12$. Furthermore, the flat coordinate $t$ and the generating function (\ref{HolGij}) of the Gromov-Witten invariants associated with $J^2$ can be also extracted as
\begin{equation}
2\pi i t=\log z,\ \
G(t)=\frac14t^2+\Delta_0(q),\ \ q=e^{2\pi i t}.
\end{equation}
This result coincides with the computation in \cite{Klemm:2007in}. With the above choice for $h(z)$, the consistent inverse intersection matrix $\eta^{-1}=2$ for a basis $J^2$ can be realized. This result provides a nontrivial verification of our conjecture (\ref{Kahl4}).

\subsubsection{Local ${\IP}^1\times{\IP}^1$: ${\cal O}(-1,-1)\oplus{\cal O}(-1,-1)\to {\IP}^1\times{\IP}^1$}

The local Calabi-Yau fourfold ${\cal O}(-1,-1)\oplus{\cal O}(-1,-1)\to {\IP}^1\times{\IP}^1$ is defined by two charge vectors $Q^1=(-1,-1,1,1,0,0)$ and $Q^2=(-1,-1,0,0,1,1)$. We denote the K\"ahler forms defined on the base ${\IP}^1\times{\IP}^1$ by $J_1$ and $J_2$. In this case, the GLSM partition function (\ref{GLSMp}) becomes
\begin{eqnarray}
Z_{\mbox{\scriptsize GLSM}}&=&
\sum_{m_1,m_2\in{\IZ}}e^{-i(\theta_1m_1+\theta_2m_2)}\int_{-\infty}^{\infty}\frac{d\sigma_1}{2\pi}\frac{d\sigma_2}{2\pi}e^{-4\pi i(r_1\sigma_1+r_2\sigma_2)}\nonumber\\
&&
\times\frac{\Gamma(-i\sigma_1-\frac{1}{2}m_1)^2}{\Gamma(1+i\sigma_1-\frac{1}{2}m_1)^2}\frac{\Gamma(-i\sigma_2-\frac{1}{2}m_2)^2}{\Gamma(1+i\sigma_2-\frac{1}{2}m_2)^2}\frac{\Gamma\big(\mathfrak{q}+i(\sigma_1+\sigma_2)+\frac12(m_1+m_2)\big)}{\Gamma\big(1-\mathfrak{q}-i(\sigma_1+\sigma_2)+\frac12(m_1+m_2)\big)}\nonumber\\
&=&
\oint\frac{d\epsilon_1}{2\pi i}\frac{d\epsilon_2}{2\pi i}(z_1\overline{z}_1)^{-\epsilon_1}(z_2\overline{z}_2)^{-\epsilon_2}
\frac{\pi^2\sin^2\pi\big(\mathfrak{q}-(\epsilon_1+\epsilon_2)\big)}{\sin^2(\pi \epsilon_1)\sin^2(\pi \epsilon_2)}\nonumber\\
&&\hspace{9em}
\times\bigg|\sum_{k_1,k_2=0}^{\infty}z_1^{k_1}z_2^{k_2}\frac{\Gamma(\mathfrak{q}+(k_1+k_2)-(\epsilon_1+\epsilon_2))^2}{\Gamma(1+k_1-\epsilon_1)^2\Gamma(1+k_2-\epsilon_2)^2}\bigg|^2,
\end{eqnarray}
where $z_{\ell}=e^{-2\pi r_{\ell}+i\theta_{\ell}}$, and the complex conjugation does not act on $\epsilon_{1,2}$. First we normalize the partition function as $\widetilde{Z}_{\mbox{\scriptsize GLSM}}=\frac{g(z_1,z_2)\overline{g(z_1,z_2)}}{f(z_1,z_2)\overline{f(z_1,z_2)}}Z_{\mbox{\scriptsize GLSM}}$, where
\begin{eqnarray}
f(z_1,z_2)&=&
\sum_{k_1,k_2=0}^{\infty}\frac{\Gamma(\mathfrak{q}+(k_1+k_2))^2}{\Gamma(1+k_1)^2\Gamma(1+k_2)^2}z_1^{k_1}z_2^{k_2},\hspace{15.5em}\\
g(z_1,z_2)&=&
\widetilde{\Gamma}(\mathfrak{q}, h(z))^2
\bigg[1+\mathfrak{q}^2\mathop{\sum_{k_1,k_2=0}^{\infty}}\limits_{(k_1,k_2)\ne (0,0)}\frac{z_1^{k_1}z_2^{k_2}}{\Gamma(1+k_1)^2\Gamma(1+k_2)^2}\prod_{j=1}^{k_1+k_2-1}(\mathfrak{q}+j)^2\bigg].
\end{eqnarray}
$h(z)\equiv h(z_1,z_2)$ is a holomorphic function of $z_{1,2}$. Taking the non-compact limit $\mathfrak{q}\to 0^+$, the partition function is expanded as
\begin{align}
\widetilde{Z}_{\mbox{\scriptsize GLSM}}=& \ 
6\mathfrak{q}^{-4}-2\big(6h(z)+6\overline{h(z)}+\log z_1z_2\overline{z}_1\overline{z}_2\big)\mathfrak{q}^{-3}\nonumber\\
&
+\big[12\big(h(z)+\overline{h(z)}\big)^2+4\big(h(z)+\overline{h(z)}\big)\log z_1z_2\overline{z}_1\overline{z}_2+\log z_1\overline{z}_1\log z_2\overline{z}_2\big]\mathfrak{q}^{-2}\nonumber\\
&
-2\big(h(z)+\overline{h(z)}\big)\big(2h(z)+2\overline{h(z)}+\log z_1\overline{z}_1\big)\big(2h(z)+2\overline{h(z)}+\log z_2\overline{z}_2\big)\mathfrak{q}^{-1}\nonumber\\
&
+\widetilde{Z}_0+{\cal O}(\mathfrak{q}),
\end{align}
where
\begin{align}
\widetilde{Z}_0=& \ 
4\big(h(z)+\overline{h(z)}\big)^4+\frac83\big(h(z)+\overline{h(z)}\big)^3\log z_1\overline{z}_1z_2\overline{z}_2+2\big(h(z)+\overline{h(z)}\big)^2\log z_1\overline{z}_1\log z_2\overline{z}_2\nonumber\\
&
+\big(\Delta_{00}(z)+\overline{\Delta_{00}(z)}\big)\log z_1\overline{z}_1\log z_2\overline{z}_2
-\frac43\zeta(3)\log z_1\overline{z}_1z_2\overline{z}_2\nonumber\\
&
+\big(\Delta_{01}(z)+\overline{\Delta_{01}(z)}\big)\log z_1\overline{z}_1
+\big(\Delta_{10}(z)+\overline{\Delta_{10}(z)}\big)\log z_2\overline{z}_2\nonumber\\
&
+2\Delta_{00}(z)\overline{\Delta_{00}(z)}+\Delta_{11}(z)+\overline{\Delta_{11}(z)}-\frac13\pi^2\big(\Delta_{00}(z)+\overline{\Delta_{00}(z)}\big).
\label{localA2Zq}
\end{align}
In the above, we have defined $\Delta_{00}(z)\equiv\Delta_{00}(z_1,z_2)$, $\Delta_{10}(z)\equiv\Delta_{10}(z_1,z_2)$, $\Delta_{01}(z)\equiv\Delta_{01}(z_1,z_2)=\Delta_{10}(z_2,z_1)$, and $\Delta_{11}(z)\equiv\Delta_{11}(z_1,z_2)$ as
\begin{align}
\Delta_{00}(z)&=\mathop{\sum_{k_1,k_2=0}^{\infty}}\limits_{(k_1,k_2)\ne (0,0)}
\frac{1}{(k_1+k_2)^2}\binom {k_1+k_2}{k_1}^2z_1^{k_1}z_2^{k_2}, \\
\Delta_{10}(z)&=2\mathop{\sum_{k_1,k_2=0}^{\infty}}\limits_{(k_1,k_2)\ne (0,0)}\frac{1}{(k_1+k_2)^2}\binom {k_1+k_2}{k_1}^2z_1^{k_1}z_2^{k_2}\big[\Psi(k_1+k_2)-\Psi(1+k_1)\big], \\
\Delta_{11}(z)&=2\mathop{\sum_{k_1,k_2=0}^{\infty}}\limits_{(k_1,k_2)\ne (0,0)}\frac{1}{(k_1+k_2)^2}\binom {k_1+k_2}{k_1}^2z_1^{k_1}z_2^{k_2}\Big[\Psi^{(1)}(k_1+k_2)
\nonumber\\
&\hspace{6em}
+2\big[\Psi(k_1+k_2)-\Psi(1+k_1)\big]\big[\Psi(k_1+k_2)-\Psi(1+k_2)\big]\Big].
\end{align}
From the comparison of (\ref{localA2Zq}) with (\ref{Kahl4}), we find that the flat coordinates $t_{1,2}$, and the generating function (\ref{HolGij}) of the Gromov-Witten invariants take the form
\begin{eqnarray}
&&
2\pi it_1=\log z_1,\ \ 2\pi it_2=\log z_2,\\
&&
\widehat{G}_{11}(t)=\widehat{G}_{22}(t)=0,\ \ \widehat{G}_{12}(t)=\Delta_{00}(q_1,q_2),\ q_{\ell}=e^{2\pi it_{\ell}}.
\end{eqnarray}
This result agrees with the computations of \cite{Klemm:2007in}.

Here let us take a basis $\{H_1,H_2,H_3\}=\{J_1^2, J_1\wedge J_2, J_2^2\}$, and consider the associated generating functions (\ref{HolFnpr}). Then, we can read off $\eta^{22}=2$ from (\ref{Kahl4}). In order to obtain the ``classical quadruple intersection numbers'' consistent with this assignment, we need to choose $h(z)=-\frac14\log (z_1z_2)$. With this choice, we obtain $\kappa_{1111}=\kappa_{2222}=-\frac58$, $\kappa_{1112}=\kappa_{1222}=\frac18$, and $\kappa_{1122}=\frac38$.

\subsubsection{Local ${\IP}^3$: ${\cal O}(-4)\to {\IP}^3$}

Finally we consider the local Calabi-Yau fourfold ${\cal O}(-4)\to {\IP}^3$ defined by the charge vector $Q=(-4,1,1,1,1)$. Let $J$ be the K\"ahler form defined on the base ${\IP}^3$. The two sphere partition function of the corresponding GLSM is given by
\begin{eqnarray}
Z_{\mbox{\scriptsize GLSM}}&=&
\sum_{m\in{\IZ}}e^{-i\theta m}\int_{-\infty}^{\infty}\frac{d\sigma}{2\pi}e^{-4\pi ir\sigma}
\frac{\Gamma(-i\sigma-\frac{1}{2}m)^4}{\Gamma(1+i\sigma-\frac{1}{2}m)^4}\frac{\Gamma(\mathfrak{q}+4i\sigma+2m)}{\Gamma(1-\mathfrak{q}-4i\sigma+2m)}\nonumber\\
&=&
\oint\frac{d\epsilon}{2\pi i}(z\overline{z})^{-\epsilon}
\frac{\pi^3\sin\pi(\mathfrak{q}-4\epsilon)}{\sin^4(\pi \epsilon)}
\bigg|\sum_{k=0}^{\infty}z^{k}\frac{\Gamma(\mathfrak{q}+4k-4\epsilon)}{\Gamma(1+k-\epsilon)^4}\bigg|^2,
\end{eqnarray}
where $z=e^{-2\pi r+i\theta}$ and the complex conjugation does not act on $\epsilon$. Then, we normalize the partition function as $\widetilde{Z}_{\mbox{\scriptsize GLSM}}=\frac{g(z)\overline{g(z)}}{f(z)\overline{f(z)}}Z_{\mbox{\scriptsize GLSM}}$, where
\begin{eqnarray}
f(z)&=&
\sum_{k=0}^{\infty}\frac{\Gamma(\mathfrak{q}+4k)}{\Gamma(1+k)^4}z^{k},\\
g(z)&=&
\widetilde{\Gamma}(\mathfrak{q}, h(z))
\bigg[1+\mathfrak{q}\sum_{k=1}^{\infty}\frac{z^k}{\Gamma(1+k)^4}\prod_{j=1}^{4k-1}(\mathfrak{q}+j)\bigg].
\end{eqnarray}
By taking the non-compact limit $\mathfrak{q}\to 0^+$, we obtain the following expansion
\begin{eqnarray}
\widetilde{Z}_{\mbox{\scriptsize GLSM}}&=&
64\mathfrak{q}^{-4}-16\big(4h(z)+4\overline{h(z)}+\log z\overline{z}\big)\mathfrak{q}^{-3}+2\big(4h(z)+4\overline{h(z)}+\log z\overline{z}\big)^2\mathfrak{q}^{-2}\nonumber\\
&&
-\Big[\frac16\big(4h(z)+4\overline{h(z)}+\log z\overline{z}\big)^3+\frac83\zeta(3)\Big]\mathfrak{q}^{-1}+\widetilde{Z}_0+{\cal O}(\mathfrak{q}).
\end{eqnarray}
Here we choose $h(z)=-\frac14 \log z$. By comparing $\widetilde{Z}_0$ with (\ref{Kahl4}), we obtain the ``consistent quadruple intersection number'' $\kappa=-\frac14$ along with the inverse intersection matrix $\eta^{-1}=-4$ for a basis $J^2$. We can also find that the topological invariant defined in (\ref{topc3j}) is given by $C=5$, and the flat coordinate $t$ and the Gromov-Witten invariants $n_d$ associated with $J^2$ are determined as
\begin{eqnarray}
&&
2\pi i t=\log z+4\sum_{k=1}^{\infty}\frac{(4k-1)!}{(k!)^4}z^k=\log z+
24z+1260z^2+123200z^3+\cdots,\\
&&
n_d=-20,~-820,~-68060,~-7486440,~-965038900,\ldots,
\end{eqnarray}
which completely agree with the result of \cite{Klemm:2007in}.

\section{Conclusion and discussions}

In this paper we have studied quantum nature of the K\"ahler moduli space of Calabi-Yau fourfolds. We utilized the recently proposed method which relates the exact two sphere partition function of an ${\cal N}=(2,2)$ GLSM to the K\"ahler potential on the quantum K\"ahler moduli space of a Calabi-Yau manifold. Especially we conjectured the explicit formula of the quantum-corrected K\"ahler potential for Calabi-Yau fourfolds. We also checked our conjecture by computing the genus zero Gromov-Witten invariants and comparing the results with mirror symmetry predictions. Since the GLSM calculation for the K\"ahler potential is reminiscent of the well-studied abelian mirror symmetry and is also applicable to non-abelian GLSMs, this method would give a clue to understand the non-abelian mirror symmetry.

Moreover, we proposed the local toric analogue of the correspondence between the GLSM partition function and the exact K\"ahler potential by extending the argument of \cite{Park:2012nn}. We also studied the exact GLSM partition functions for local toric Calabi-Yau varieties in a similar manner to the cases of compact Calabi-Yau manifold. In order to realize the expected classical terms (intersection numbers) of the K\"ahler potential for a local toric Calabi-Yau variety, we need to modify the normalization of the corresponding GLSM partition function. In this normalization, we have introduced a holomorphic function $h(z)$ which should take the form $h(z)=\sum_{\ell}c_{\ell}\log z_{\ell}$. For local toric Calabi-Yau fourfolds, we fixed the constants $c_{\ell}$ from the consistency with the intersection matrix on $H^{2,2}_{\mbox{\scriptsize prim}}$ appeared in our conjectural formula for the K\"ahler potential.

An immediate generalization of our work is to find the exact K\"ahler potential for higher dimensional Calabi-Yau manifolds with $d \ge 5$. Once the explicit formula is found, the GLSM calculation would provide an efficient way to compute the Gromov-Witten invariants of Calabi-Yau manifold with arbitrary dimension.

As demonstrated in \cite{Sharpe:2012ji}, the exact GLSM partition function is also useful to study the Landau-Ginzburg phase of GLSM which describes a Calabi-Yau manifold at the large radius point. It would be also interesting to study such a phase transition using the GLSM partition function for not only three dimension but also higher dimensions.

Our conjecture about the exact K\"ahler potential for Calabi-Yau fourfold allows one to study the nonperturbative aspects of the F-theory compactification. In contrast to the Type IIB string compactification, corrections to the tree level K\"ahler potential has yet to be fully understood in the F-theory compactification. We hope that many applications of our result would reveal themselves.

\subsection*{Acknowledgements}

We would like to thank Rajesh Gopakumar and Anshuman Maharana for useful discussions and comments. We would also like to thank the referee for several comments and remarks.

\appendix
\section{Note on Grassmannian Calabi-Yau manifold}

In this appendix, we first review the computation of the classical cohomology ring of the Grassmannian $G(k,n)$. Then we consider the GLSM description of general complete intersection Calabi-Yau manifold in $G(k,n)$ and generalize the relation between (\ref{GLSMGrcoeff}) and (\ref{Grass4Peri}). We also summarize some computational results for topological invariants of Grassmannian Calabi-Yau fourfold.

\subsection{Schubert calculus and Chern classes of Grassmannian}

The Grassmannian $G(k,n)$ is defined by the set of $k$-planes $\Lambda$ in ${\IC}^n$. Bases of $\Lambda$ consist of $kn$ components up to the $GL(k,{\IC})$ action on $\Lambda$. Thus the dimension of $G(k,n)$ is given by $k(n-k)$. The cohomology ring of $G(k,n)$ is described by the classes of the Schubert cycles
\begin{equation}
\sigma_{a_1,\ldots,a_k}(V)=\big\{\Lambda\in G(k,n)\big|\dim (\Lambda\cap V_{n-k+i-a_i})\ge i,\ i=1,\ldots,k \big\},
\end{equation}
which generate the integral homology (see, for example, \cite{Griffiths:1978}). Here $V=(V_1\subset V_2\subset \cdots \subset V_n)$ is a flag composed by $i$-dimensional subspaces $V_i$ in ${\IC}^n$ and the index $a_i$ is an integer sequence satisfying $n-k\ge a_1 \ge a_2 \ge \cdots \ge a_k \ge 0$.
The codimension of Schubert cycle $\sigma_{\vec{a}}\equiv\sigma_{a_1,\ldots,a_k}(V)$ is given by $|\vec{a}|\equiv\sum_{i=1}^ka_i$.

The intersection number of Schubert cycles can be computed by Pieri's formula
\begin{equation}
\sigma_{a}\cdot \sigma_{\vec{b}}=\mathop{\sum_{b_i\le c_i\le b_{i-1}}}\limits_{|\vec{c}|=a+|\vec{b}|}\sigma_{\vec{c}},
\end{equation}
where $\sigma_a \equiv \sigma_{a,0,\ldots,0}(V)$ is called the special Schubert cycle. Because any Schubert cycles $\sigma_{a_1,\ldots,a_k}$ can be represented in terms of the special Schubert cycles by Giambelli's formula
\begin{equation}
\sigma_{a_1,\ldots,a_k}=
\left|\begin{array}{cccc}
\sigma_{a_1} & \sigma_{a_1+1} & \cdots & \sigma_{a_1+k-1}\\
\sigma_{a_2-1} & \sigma_{a_2} & \cdots & \sigma_{a_2+k-2}\\
\vdots & \vdots & \ddots & \vdots\\
\sigma_{a_k-k+1} & \sigma_{a_k-k+2} & \cdots & \sigma_{a_k}
\end{array}\right|,
\end{equation}
the cohomology ring of $G(k,n)$ is generated by the classes of the special Schubert cycles \cite{Griffiths:1978}. In the following we summarize some results of the intersection numbers of the special Schubert cycles. Note that, in order to define the intersection number of Schubert cycles $\sigma_{\vec{a}_{\ell}}$, $\ell=1,\ldots,p$, we need to require that the codimension of the union of those cycles is equal to the dimension of $G(k,n)$, i.e. $\sum_{\ell=1}^p |\vec{a}_{\ell}|=k(n-k)$.

\vspace{0.5em}
\noindent\underline{$G(2,5)$}:
\begin{equation}
\sigma_1^6=\sigma_1^4(\sigma_2+\sigma_{1,1})=\sigma_1^3(\sigma_3+2\sigma_{2,1})=
\sigma_1^2(3\sigma_{3,1}+2\sigma_{2,2})=\sigma_1\cdot 5\sigma_{3,2}=5\sigma_{3,3}=5.
\end{equation}
Similarly, we can obtain
\begin{equation}
\sigma_1^4\sigma_2=3,\ \ \sigma_1^3\sigma_3=1,\ \ \sigma_1^2\sigma_2^2=2.
\end{equation}

\noindent\underline{$G(2,6)$}:
\begin{equation}
\sigma_1^8=14,\ \ \sigma_1^6\sigma_2=9,\ \ \sigma_1^5\sigma_3=4,\ \ \sigma_1^4\sigma_2^2=6,\ \ \sigma_1^4\sigma_4=1.
\end{equation}

\noindent\underline{$G(2,7)$}:
\begin{equation}
\sigma_1^{10}=42,\ \ \sigma_1^8\sigma_2=28,\ \ \sigma_1^7\sigma_3=14,\ \ \sigma_1^6\sigma_2^2=19,\ \ \sigma_1^6\sigma_4=5.
\end{equation}

\noindent\underline{$G(2,8)$}:
\begin{equation}
\sigma_1^{12}=132,\ \ \sigma_1^{10}\sigma_2=90,\ \ \sigma_1^9\sigma_3=48,\ \ \sigma_1^8\sigma_4=20,\ \ \sigma_1^8\sigma_2^2=62.
\end{equation}

\noindent\underline{$G(3,6)$}:
\begin{equation}
\sigma_1^9=42,\ \ \sigma_1^7\sigma_2=21,\ \ \sigma_1^6\sigma_3=5,\ \ \sigma_1^5\sigma_2^2=11.
\end{equation}

\vspace{0.5em}
In the following, we denote the Poincar\'e dual of the special Schubert cycle by the same symbol $\sigma_a$. The total Chern class of Grassmannian $G(k,n)$ is given by \cite{Borel:1958,Haghighat:2008ut}
\begin{equation}
c(G(k,n))=\prod_{i=1}^{n-k}(1-x_i)^n\cdot \prod_{i,j=1}^{n-k}\big(1-(x_i-x_j)^2\big)^{-\frac12},
\label{Grass_tchern}
\end{equation}
where the $p$-th Chern class $c_p(G(k,n))$ is obtained as a coefficient of $h^p$ after changing all the variables $x_{i}$ into $hx_{i}$ and taking the series expansion of $h$. Then $c_p(G(k,n))$ can be expressed in terms of the elementary symmetric polynomials $e_a(x_i)=\sum_{i_1<\cdots <i_a}x_{i_1}\cdots x_{i_a}$ which are identified with the cohomology classes $\sigma_a$. In the following, we summarize the results for several examples.

\vspace{0.5em}
\noindent\underline{$G(2,5)$}:
\begin{eqnarray}
&&
c_1(G(2,5))=5\sigma_1,\ \ c_2(G(2,5))=12\sigma_1^2-\sigma_2,\ \ c_3(G(2,5))=20\sigma_1^3-10\sigma_1\sigma_2+5\sigma_3,\nonumber\\
&&
c_4(G(2,5))=28\sigma_1^4-38\sigma_1^2\sigma_2+20\sigma_1\sigma_3+7\sigma_2^2-210\sigma_4.
\end{eqnarray}

\noindent\underline{$G(2,6)$}:
\begin{eqnarray}
&&
c_1(G(2,6))=6\sigma_1,\ \ c_2(G(2,6))=18\sigma_1^2-2\sigma_2,\ \ c_3(G(2,6))=38\sigma_1^3-18\sigma_1\sigma_2+6\sigma_3,\nonumber\\
&&
c_4(G(2,6))=66\sigma_1^4-74\sigma_1^2\sigma_2+32\sigma_1\sigma_3+9\sigma_2^2-2\sigma_4.
\end{eqnarray}

\noindent\underline{$G(2,7)$}:
\begin{eqnarray}
&&
c_1(G(2,7))=7\sigma_1,\ \ c_2(G(2,7))=25\sigma_1^2-3\sigma_2,\ \ c_3(G(2,7))=63\sigma_1^3-28\sigma_1\sigma_2+7\sigma_3,\nonumber\\
&&
c_4(G(2,7))=129\sigma_1^4-127\sigma_1^2\sigma_2+46\sigma_1\sigma_3+12\sigma_2^2-3\sigma_4.
\end{eqnarray}

\noindent\underline{$G(2,8)$}:
\begin{eqnarray}
&&
c_1(G(2,8))=8\sigma_1,\ \ c_2(G(2,8))=33\sigma_1^2-4\sigma_2,\ \ c_3(G(2,8))=96\sigma_1^3-40\sigma_1\sigma_2+8\sigma_3,\nonumber\\
&&
c_4(G(2,8))=225\sigma_1^4-200\sigma_1^2\sigma_2+62\sigma_1\sigma_3+16\sigma_2^2-4\sigma_4.
\end{eqnarray}

\noindent\underline{$G(3,6)$}:
\begin{eqnarray}
&&
c_1(G(3,6))=6\sigma_1,\ \ c_2(G(3,6))=17\sigma_1^2,\ \ c_3(G(3,6))=32\sigma_1^3-6\sigma_1\sigma_2+6\sigma_3,\nonumber\\
&&
c_4(G(3,6))=48\sigma_1^4-36\sigma_1^2\sigma_2+30\sigma_1\sigma_3+6\sigma_2^2-372\sigma_4.
\end{eqnarray}

\subsection{GLSM description for Grassmannian Calabi-Yau manifold}

\begin{table}[t]
\begin{center}
\begin{tabular}{c|c|c}
\hline
Field & $U(k)$  & $U(1)_V$ \\ \hline
$\Phi^i$ & $\mathbf{k}_{+1}$ & $2\mathfrak{q}$ \\
$P_a$ & $\mathbf{1}_{-kd_a}$ & $2-2kd_a\mathfrak{q}$ \\ \hline
\end{tabular}
\caption{Matter content of the $U(k)$ GLSM describing the Grassmannian Calabi-Yau $d$-fold $X_{d_1,\ldots,d_r}\subset G(k,n)$. Here $i=1,\ldots,n$ and $a=1,\ldots,r$. The subscript denotes the charge under the central $U(1)\subset U(k)$.}
\label{grassgen}
\end{center}
\end{table}
Let us consider a $d$ dimensional Calabi-Yau manifold $X_{d_1,\ldots,d_r}$ defined by a complete intersection of $r$ hyperplanes with degrees $(d_1,\ldots,d_r)$ in the Grassmannian $G(k,n)$. The complex dimension of $X_{d_1,\ldots,d_r}$ is given by $d=kn-k^2-r$ and the Calabi-Yau condition $d_1+\cdots +d_r=n$ must be satisfied. The total Chern class of this manifold is \cite{Haghighat:2008ut}
\begin{equation}
c(X_{d_1,\ldots,d_r})=\frac{c(G(k,n))}{(1+d_1\sigma_1)\cdots (1+d_r\sigma_1)}.
\end{equation}
Using this formula and (\ref{Grass_tchern}), one can compute the classical topological invariants. For example, the intersection number and the Euler characteristic are given by
\begin{eqnarray}
&&
\kappa=\int_{X_{d_1,\ldots,d_r}}\sigma_1^d=\int_{G(k,n)}\sigma_1^d\wedge \prod_{a=1}^rd_a\sigma_1,\\
&&
\chi=\int_{X_{d_1,\ldots,d_r}}c_d(X_{d_1,\ldots,d_r})=\int_{G(k,n)}c_d(X_{d_1,\ldots,d_r})\wedge \prod_{a=1}^rd_a\sigma_1.
\end{eqnarray}

The Grassmannian Calabi-Yau $d$-fold $X_{d_1,\ldots,d_r} \subset G(k,n)$ can be described by the $U(k)$ GLSM with matter multiplets shown in Table \ref{grassgen}. The superpotential is given by $W=\overset{r}{\underset{{a=1}}{\sum}}
P_aW_{d_a}(B)$, where $W_{d_a}(B)$ is a degree $d_a$ polynomial in the baryonic variables\footnote{These variables are called Pl\"ucker coordinates corresponding to the homogeneous coordinates on the projective space in which the Grassmannian is embedded.} $B^{i_1\ldots i_k}=\epsilon^{I_1\ldots I_k}\Phi^{i_1}_{I_1}\cdots \Phi^{i_k}_{I_k}$ \cite{Hori:2006dk}. Then we can compute the two sphere partition function (\ref{GLSMp}) in the same way as in Section 4.4. In the Grassmann phase $r\gg 0$, we obtain
\begin{align}
&
Z_{\mbox{\scriptsize GLSM}}=
(-1)^{\frac12k(k-1)}\frac{1}{k!}(z\overline{z})^{k\mathfrak{q}}\oint\frac{d\epsilon_1\cdots d\epsilon_k}{(2\pi i)^k}(z\overline{z})^{-\sum_{i=1}^k\epsilon_i}
\frac{\pi^{kn-r}\prod_{a=1}^r\sin\big(\pi d_a\sum_{i=1}^k\epsilon_i\big)}{\prod_{i=1}^k\sin^n(\pi \epsilon_i)}\nonumber\\
&\hspace{1em}
\times\bigg|\sum_{\ell_1,\ldots \ell_k=0}^{\infty}\big((-1)^nz\big)^{\sum_{i=1}^k\ell_i}\prod_{1\le i<j \le k}\big[(\ell_i-\ell_j)-(\epsilon_i-\epsilon_j)\big]\cdot\frac{\prod_{a=1}^r\Gamma\big(1+d_a\sum_{i=1}^k(\ell_i-\epsilon_i)\big)}{\prod_{i=1}^k\Gamma(1+\ell_i-\epsilon_i)^n}\bigg|^2,
\label{Gknpart}
\end{align}
where $z=e^{-2\pi r+i\theta}$.

Here we consider the case of $k=2$. As demonstrated in Section 4, we can extract a holomorphic function
\begin{eqnarray}
&&
T^0(z)=\sum_{k_1,k_2=0}^{\infty}(-z)^{k_1+k_2}\frac{(d_1(k_1+k_2))!\cdots (d_r(k_1+k_2))!}{(k_1!k_2!)^n}\nonumber\\
&&\hspace{8em}
\times\Big[1-\frac{n}{2}(k_1-k_2)\big(\Psi(1+k_1)-\Psi(1+k_2)\big)\Big],
\end{eqnarray}
which gives a normalization of the partition function (\ref{Gknpart}). On the other hand, the fundamental period of the corresponding mirror manifold is given by \cite{Batyrev:1998kx,Batyrev:1998mit}
\begin{equation}
\widehat{T}^0(z)=\sum_{\ell_0,\ell_1,\ldots,\ell_{n-3}=0}^{\infty}z^{\ell_0}\frac{(d_1\ell_0)!\cdots (d_r\ell_0)!}{(\ell_0!)^n}\prod_{i=1}^{n-3}
\binom {\ell_0}{\ell_i}\binom {\ell_{i+1}}{\ell_i},
\end{equation}
where we defined $\ell_{n-2}\equiv\ell_0$. We can check the coincidence between $T^0(z)$ and $\widehat{T}^0(z)$ up to higher order in $z$ and thus the identity
\begin{equation}
T^0(z)=\widehat{T}^0(z)
\label{Grass_genPeri}
\end{equation}
is expected to hold exactly for $d_a\ge 0$ and $n\ge 3$. Note that the Calabi-Yau condition $d_1+\cdots +d_r=n$ need not be satisfied for this identity. Generalizing this identity to arbitrary $k$ is straightforward. It will be interesting to prove this identity from the perspective of combinatorics.

\subsection{Gromov-Witten invariants of Grassmannian fourfolds}

In $d=4$, we can list all the Grassmannian Calabi-Yau manifold $X_{d_1,\ldots,d_r}\subset G(k,n)$ as
\begin{eqnarray}
&&
X_{1,4}\subset G(2,5),\ \ X_{2,3}\subset G(2,5),\ \ X_{1^3,3}\subset G(2,6),\nonumber\\
&&
X_{1^2,2^2}\subset G(2,6),\ \ X_{1^5,2}\subset G(2,7),\ \ X_{1^4,2}\subset G(3,6),
\label{Grass_list}
\end{eqnarray}
including $X_{1^8}\subset G(2,8)$ discussed in Section 4.4. We checked that our conjecture (\ref{Kahl4}) holds in these cases, and the Gromov-Witten invariants associated with $H_1\equiv\sigma_1^2$ are summarised in Table \ref{grass1GW}.

For each Calabi-Yau fourfold presented in (\ref{Grass_list}), one finds that $\dim H^{2,2}_{\mbox{\scriptsize prim}}=2$. Just like in Section 4.4, by taking an element $H_2$ orthogonal to $H_1$, we obtain the generating function (\ref{HolFnpr}) of the corresponding Gromov-Witten invariants. The results are summarized in Table \ref{grass2GW}. The Hodge numbers\footnote{Method used in calculating these is explained e.g. in \cite{Klemm:1996ts}.}, some classical topological invariants, an element $H_2$ which we have taken, the intersection matrix $\eta_{mn}$ in the basis $\{H_1,H_2\}$, and the Picard-Fuchs operator ${\cal D}$ are listed below. For $X_{1^4,2}\subset G(3,6)$, we have found that there are no worldsheet instanton corrections to the correlator (\ref{correl1124}) associated with $H_2=2\sigma_2-\sigma_1^2=\sigma_2-\sigma_{1,1}$.
\begin{table}[t]
\begin{center}
\begin{tabular}{c|r|r|r}
\hline
$$ & $X_{1,4}\subset G(2,5)$ & $X_{2,3}\subset G(2,5)$ & $X_{1^3,3}\subset G(2,6)$ \\ \hline
$n_1$ & 9440 & 5580 & 4158 \\
$n_2$ & 4383680 & 1102770 & 538272 \\
$n_3$ & 3701308960 & 391989240 & 115394706 \\
$n_4$ & 4126541676160 & 183418036920 & 32820139926 \\
$n_5$ & 5368332901844000 & 100068916666500 & 10856106949968 \\
\hline
$$ & $X_{1^2,2^2}\subset G(2,6)$ & $X_{1^5,2}\subset G(2,7)$ & $X_{1^4,2}\subset G(3,6)$ \\ \hline
$n_1$ & 3136 & 2296 & 2520 \\
$n_2$ & 242032 & 112196 & 112140 \\
$n_3$ & 30787008 & 8076880 & 8494920 \\
$n_4$ & 5179177248 & 781233880 & 829679760 \\
$n_5$ & 1012577938176 & 87311729064 & 94209368400 \\
\hline
\end{tabular}
\caption{Gromov-Witten invariants associated with $H_1=\sigma_1^2$.}
\label{grass1GW}
\end{center}
\end{table}
\begin{table}[t]
\begin{center}
\begin{tabular}{c|r|r|r|r|r}
\hline
$$ & $X_{1,4}\subset G(2,5)$ & $X_{2,3}\subset G(2,5)$ & $X_{1^3,3}\subset G(2,6)$ & $X_{1^2,2^2}\subset G(2,6)$ & $X_{1^5,2}\subset G(2,7)$ \\ \hline
$m_1$ & 480 & 360 & 756 & 672 & 112 \\
$m_2$ & 72960 & 21240 & 23058 & 11424 & 980  \\
$m_3$ & 63993120 & 7998480 & 5788692 & 1731744 & 89152  \\
$m_4$ & 68283664320 & 3580395840 & 1555295364 & 275167200 & 8067556  \\
$m_5$ & 86760415092000 & 1907243811000 & 503261635464 & 52620757056 & 884735376  \\
\hline
\end{tabular}
\caption{Gromov-Witten invariants associated with $H_2$ orthogonal to $H_1$.}
\label{grass2GW}
\end{center}
\end{table}

\vspace{0.5em}
\noindent\underline{$X_{1,4}\subset G(2,5)$}:
\begin{eqnarray}
&&
h^{1,1}=1,\ \ h^{2,1}=0,\ \ h^{2,2}=1244,\ \ h^{3,1}=299,\\
&&
\chi=1848,\ \ \int_{X_{1,4}}c_3\wedge \sigma_1=-440,\ \ \int_{X_{1,4}}c_2\wedge \sigma_1^2=148,\ \ \kappa=20,\\
&&
H_2=5\sigma_2-3\sigma_1^2,\ \ \eta_{mn}=\rm{diag}(20,\ 20),\\
&&
{\cal D}=
(\Theta-1)\Theta^5-8z(4\Theta+3)(4\Theta+1)(2\Theta+1)(11\Theta^2+11\Theta+3)\Theta\nonumber\\
&&\hspace{2em}
-64z^2(4\Theta+7)(4\Theta+5)(4\Theta+3)(4\Theta+1)(2\Theta+3)(2\Theta+1).
\end{eqnarray}

\noindent\underline{$X_{2,3}\subset G(2,5)$}:
\begin{eqnarray}
&&
h^{1,1}=1,\ \ h^{2,1}=0,\ \ h^{2,2}=804,\ \ h^{3,1}=189,\\
&&
\chi=1188,\ \ \int_{X_{2,3}}c_3\wedge \sigma_1=-360,\ \ \int_{X_{2,3}}c_2\wedge \sigma_1^2=162,\ \ \kappa=30,\\
&&
H_2=5\sigma_2-3\sigma_1^2,\ \ \eta_{mn}=\rm{diag}(30,\ 30),\\
&&
{\cal D}=
(\Theta-1)\Theta^5-6z(3\Theta+2)(3\Theta+1)(2\Theta+1)(11\Theta^2+11\Theta+3)\Theta\nonumber\\
&&\hspace{2em}
-36z^2(3\Theta+5)(3\Theta+4)(3\Theta+2)(3\Theta+1)(2\Theta+3)(2\Theta+1).
\end{eqnarray}

\noindent\underline{$X_{1^3,3}\subset G(2,6)$}:
\begin{eqnarray}
&&
h^{1,1}=1,\ \ h^{2,1}=0,\ \ h^{2,2}=924,\ \ h^{3,1}=219,\\
&&
\chi=1368,\ \ \int_{X_{1^3,3}}c_3\wedge \sigma_1=-426,\ \ \int_{X_{1^3,3}}c_2\wedge \sigma_1^2=198,\ \ \kappa=42,\\
&&
H_2=14\sigma_2-9\sigma_1^2,\ \ \eta_{mn}=\rm{diag}(42,\ 126),\\
&&
{\cal D}=
(\Theta-1)\Theta^5-3z(3\Theta+2)(3\Theta+1)(2\Theta+1)(13\Theta^2+13\Theta+4)\Theta\nonumber\\
&&\hspace{2em}
-27z^2(3\Theta+5)(3\Theta+4)^2(3\Theta+2)^2(3\Theta+1).
\end{eqnarray}

\noindent\underline{$X_{1^2,2^2}\subset G(2,6)$}:
\begin{eqnarray}
&&
h^{1,1}=1,\ \ h^{2,1}=0,\ \ h^{2,2}=604,\ \ h^{3,1}=139,\\
&&
\chi=888,\ \ \int_{X_{1^2,2^2}}c_3\wedge \sigma_1=-344,\ \ \int_{X_{1^2,2^2}}c_2\wedge \sigma_1^2=208,\ \ \kappa=56,\\
&&
H_2=14\sigma_2-9\sigma_1^2,\ \ \eta_{mn}=\rm{diag}(56,\ 168),\\
&&
{\cal D}=
(\Theta-1)\Theta^5-4z(2\Theta+1)^3(13\Theta^2+13\Theta+4)\Theta\nonumber\\
&&\hspace{2em}
-48z^2(3\Theta+4)(3\Theta+2)(2\Theta+3)^2(2\Theta+1)^2.
\end{eqnarray}

\noindent\underline{$X_{1^5,2}\subset G(2,7)$}:
\begin{eqnarray}
&&
h^{1,1}=1,\ \ h^{2,1}=0,\ \ h^{2,2}=576,\ \ h^{3,1}=132,\\
&&
\chi=846,\ \ \int_{X_{1^5,2}}c_3\wedge \sigma_1=-364,\ \ \int_{X_{1^5,2}}c_2\wedge \sigma_1^2=252,\ \ \kappa=84,\\
&&
H_2=3\sigma_2-2\sigma_1^2,\ \ \eta_{mn}=\rm{diag}(84,\ 6),\\
&&
{\cal D}=
9(\Theta-1)\Theta^5-6z(310\Theta^5+919\Theta^4+884\Theta^3+476\Theta^2+132\Theta+15)\Theta\nonumber\\
&&\hspace{2em}
-4z^2(21311\Theta^6+78951\Theta^5+154395\Theta^4+180544\Theta^3+121086\Theta^2+42546\Theta+6048)\nonumber\\
&&\hspace{2em}
-8z^3(2\Theta+1)(57561\Theta^5+249372\Theta^4+412273\Theta^3+310581\Theta^2+104388\Theta+11691)\nonumber\\
&&\hspace{2em}
-16z^4(2\Theta+3)(2\Theta+1)(10501\Theta^4+20138\Theta^3+13096\Theta^2+2676\Theta-154)\nonumber\\
&&\hspace{2em}
+1184z^5(2\Theta+5)(2\Theta+3)(2\Theta+1)(\Theta+1)^3.
\end{eqnarray}

\noindent\underline{$X_{1^8}\subset G(2,8)$}:
\begin{eqnarray}
&&
h^{1,1}=1,\ \ h^{2,1}=0,\ \ h^{2,2}=436,\ \ h^{3,1}=97,\\
&&
\chi=636,\ \ \int_{X_{1^8}}c_3\wedge \sigma_1=-336,\ \ \int_{X_{1^8}}c_2\wedge \sigma_1^2=300,\ \ \kappa=132.
\end{eqnarray}

\noindent\underline{$X_{1^4,2}\subset G(3,6)$}:
\begin{eqnarray}
&&
h^{1,1}=1,\ \ h^{2,1}=0,\ \ h^{2,2}=564,\ \ h^{3,1}=129,\\
&&
\chi=828,\ \ \int_{X_{1^4,2}}c_3\wedge \sigma_1=-360,\ \ \int_{X_{1^4,2}}c_2\wedge \sigma_1^2=252,\ \ \kappa=84,\\
&&
{\cal D}=
\Theta^5-2z(2\Theta+1)(65\Theta^4+130\Theta^3+105\Theta^2+40\Theta+6)\nonumber\\
&&\hspace{2em}
+16z^2(4\Theta+5)(4\Theta+3)(2\Theta+3)(2\Theta+1)(\Theta+1).
\end{eqnarray}



\begin{thebibliography}{99}
\parskip=-2pt

\bibitem{Dixon:1989fj} 
  L.~J.~Dixon, V.~Kaplunovsky and J.~Louis,
  ``On Effective Field Theories Describing $(2,2)$ Vacua of the Heterotic String,''
  Nucl.\ Phys.\ B {\bf 329}, 27 (1990).

\bibitem{Candelas:1990rm} 
  P.~Candelas, X.~C.~De La Ossa, P.~S.~Green and L.~Parkes,
  ``A Pair of Calabi-Yau manifolds as an exactly soluble superconformal theory,''
  Nucl.\ Phys.\ B {\bf 359}, 21 (1991).

\bibitem{Greene:1993vm} 
  B.~R.~Greene, D.~R.~Morrison and M.~R.~Plesser,
  ``Mirror manifolds in higher dimension,''
  Commun.\ Math.\ Phys.\  {\bf 173}, 559 (1995)
  [hep-th/9402119].

\bibitem{Mayr:1996sh} 
  P.~Mayr,
  ``Mirror symmetry, $N=1$ superpotentials and tensionless strings on Calabi-Yau four folds,''
  Nucl.\ Phys.\ B {\bf 494}, 489 (1997)
  [hep-th/9610162].

\bibitem{Klemm:1996ts} 
  A.~Klemm, B.~Lian, S.~S.~Roan and S.~-T.~Yau,
  ``Calabi-Yau fourfolds for M theory and F theory compactifications,''
  Nucl.\ Phys.\ B {\bf 518}, 515 (1998)
  [hep-th/9701023].

\bibitem{Klemm:2007in} 
  A.~Klemm and R.~Pandharipande,
  ``Enumerative geometry of Calabi-Yau 4-folds,''
  Commun.\ Math.\ Phys.\  {\bf 281}, 621 (2008)
  [arXiv:0702189 [math.AG]].

\bibitem{Witten:1993yc} 
  E.~Witten,
  ``Phases of $N=2$ theories in two-dimensions,''
  Nucl.\ Phys.\ B {\bf 403}, 159 (1993)
  [hep-th/9301042].

\bibitem{Hosono:1993qy} 
  S.~Hosono, A.~Klemm, S.~Theisen and S.~-T.~Yau,
  ``Mirror symmetry, mirror map and applications to Calabi-Yau hypersurfaces,''
  Commun.\ Math.\ Phys.\  {\bf 167}, 301 (1995)
  [hep-th/9308122].

\bibitem{Hosono:1994ax} 
  S.~Hosono, A.~Klemm, S.~Theisen and S.~-T.~Yau,
  ``Mirror symmetry, mirror map and applications to complete intersection Calabi-Yau spaces,''
  Nucl.\ Phys.\ B {\bf 433}, 501 (1995)
  [hep-th/9406055].

\bibitem{Hori:2000kt} 
  K.~Hori and C.~Vafa,
  ``Mirror symmetry,''
  hep-th/0002222.

\bibitem{Jockers:2012dk} 
  H.~Jockers, V.~Kumar, J.~M.~Lapan, D.~R.~Morrison and M.~Romo,
  ``Two-Sphere Partition Functions and Gromov-Witten Invariants,''
  arXiv:1208.6244 [hep-th].

\bibitem{Benini:2012ui} 
  F.~Benini and S.~Cremonesi,
  ``Partition functions of ${\cal N}=(2,2)$ gauge theories on $S^2$ and vortices,''
  arXiv:1206.2356 [hep-th].

\bibitem{Doroud:2012xw} 
  N.~Doroud, J.~Gomis, B.~Le Floch and S.~Lee,
  ``Exact Results in $D=2$ Supersymmetric Gauge Theories,''
  arXiv:1206.2606 [hep-th].

\bibitem{Gomis:2012wy} 
  J.~Gomis and S.~Lee,
  ``Exact Kahler Potential from Gauge Theory and Mirror Symmetry,''
  arXiv:1210.6022 [hep-th].

\bibitem{Witten:1988xj}
  E.~Witten,
  ``Topological Sigma Models,''
  Commun.\ Math.\ Phys.\  {\bf 118}, 411 (1988).

\bibitem{Witten:1991zz} 
  E.~Witten,
  ``Mirror manifolds and topological field theory,''
  In *Yau, S.T. (ed.): Mirror symmetry I* 121-160
  [hep-th/9112056].

\bibitem{Batyrev:1993dm} 
  V.~V.~Batyrev,
  ``Dual Polyhedra and Mirror Symmetry for Calabi-Yau Hypersurfaces in Toric Varieties,''
  alg-geom/9310003.

\bibitem{Batyrev:1994pg} 
  V.~V.~Batyrev and L.~A.~Borisov,
  ``On Calabi-Yau complete intersections in toric varieties,''
  alg-geom/9412017.

\bibitem{Aspinwall:1991ce} 
  P.~S.~Aspinwall and D.~R.~Morrison,
  ``Topological field theory and rational curves,''
  Commun.\ Math.\ Phys.\  {\bf 151}, 245 (1993)
  [hep-th/9110048].

\bibitem{Gopakumar:1998ii} 
  R.~Gopakumar and C.~Vafa,
  ``M theory and topological strings. 1.,''
  hep-th/9809187.

\bibitem{Gopakumar:1998jq} 
  R.~Gopakumar and C.~Vafa,
  ``M theory and topological strings. 2.,''
  hep-th/9812127.

\bibitem{Ceresole:1992su} 
  A.~Ceresole, R.~D'Auria, S.~Ferrara, W.~Lerche and J.~Louis,
  ``Picard-Fuchs equations and special geometry,''
  Int.\ J.\ Mod.\ Phys.\ A {\bf 8}, 79 (1993)
  [hep-th/9204035].

\bibitem{Strominger:1990pd} 
  A.~Strominger,
  ``Special Geometry,''
  Commun.\ Math.\ Phys.\  {\bf 133}, 163 (1990).

\bibitem{Candelas:1990pi} 
  P.~Candelas and X.~de la Ossa,
  ``Moduli Space Of Calabi-yau Manifolds,''
  Nucl.\ Phys.\ B {\bf 355}, 455 (1991).

\bibitem{Craps:1997gp} 
  B.~Craps, F.~Roose, W.~Troost and A.~Van Proeyen,
  ``What is special Kahler geometry?,''
  Nucl.\ Phys.\ B {\bf 503}, 565 (1997)
  [hep-th/9703082].

\bibitem{Ferrara:1989vp} 
  S.~Ferrara and A.~Strominger,
  ``$N=2$ Space-time Supersymmetry And Calabi-yau Moduli Space,''
  Conf.\ Proc.\ C {\bf 8903131}, 245 (1989).

\bibitem{Goddard:1976qe} 
  P.~Goddard, J.~Nuyts and D.~I.~Olive,
  ``Gauge Theories and Magnetic Charge,''
  Nucl.\ Phys.\ B {\bf 125}, 1 (1977).

\bibitem{Park:2012nn} 
  D.~S.~Park and J.~Song,
  ``The Seiberg-Witten Kahler Potential as a Two-Sphere Partition Function,''
  arXiv:1211.0019 [hep-th].

\bibitem{Grisaru:1986px} 
  M.~T.~Grisaru, A.~E.~M.~van de Ven and D.~Zanon,
  ``Four Loop beta Function for the $N=1$ and $N=2$ Supersymmetric Nonlinear Sigma Model in Two-Dimensions,''
  Phys.\ Lett.\ B {\bf 173}, 423 (1986).

\bibitem{Grimm:2009ef} 
  T.~W.~Grimm, T.~-W.~Ha, A.~Klemm and D.~Klevers,
  ``Computing Brane and Flux Superpotentials in F-theory Compactifications,''
  JHEP {\bf 1004}, 015 (2010)
  [arXiv:0909.2025 [hep-th]].

\bibitem{Alim:2009bx} 
  M.~Alim, M.~Hecht, H.~Jockers, P.~Mayr, A.~Mertens and M.~Soroush,
  ``Hints for Off-Shell Mirror Symmetry in type II/F-theory Compactifications,''
  Nucl.\ Phys.\ B {\bf 841}, 303 (2010)
  [arXiv:0909.1842 [hep-th]].

\bibitem{Alim:2011rp} 
  M.~Alim, M.~Hecht, H.~Jockers, P.~Mayr, A.~Mertens and M.~Soroush,
  ``Flat Connections in Open String Mirror Symmetry,''
  JHEP {\bf 1206}, 138 (2012)
  [arXiv:1110.6522 [hep-th]].

\bibitem{Jockers:2012zr} 
  H.~Jockers, V.~Kumar, J.~M.~Lapan, D.~R.~Morrison and M.~Romo,
  ``Nonabelian 2D Gauge Theories for Determinantal Calabi-Yau Varieties,''
  JHEP {\bf 1211}, 166 (2012)
  [arXiv:1205.3192 [hep-th]].

\bibitem{Hosono:2011np} 
  S.~Hosono and H.~Takagi,
  ``Mirror symmetry and projective geometry of Reye congruences I,''
  arXiv:1101.2746 [math.AG].

\bibitem{Hosono:2012hc} 
  S.~Hosono and H.~Takagi,
  ``Determinantal Quintics and Mirror Symmetry of Reye Congruences,''
  arXiv:1208.1813 [math.AG].

\bibitem{Li:1998hba} 
  A.~-M.~Li and Y.~Ruan,
  ``Symplectic surgery and Gromov-Witten invariants of Calabi-Yau 3-folds,''
  Invent.\ Math.\ {\bf 145}, 151-218 (2001) [arXiv:9803036 [math.AG]].

\bibitem{Batyrev:1998kx} 
  V.~V.~Batyrev, I.~Ciocan-Fontanine, B.~Kim and D.~van Straten,
  ``Conifold Transitions and Mirror Symmetry for Calabi-Yau Complete Intersections in Grassmannians,''
  Nucl.\ Phys.\ B {\bf 514}, 640 (1998) [alg-geom/9710022].

\bibitem{Sturmfels:1996} 
  B.~Sturmfels,
  ``Gr\"obner Bases and Convex Polytopes,''
  Univ.\ Lect.\ Notes, vo. 8, AMS, 1996.

\bibitem{Hori:2006dk} 
  K.~Hori and D.~Tong,
  ``Aspects of Non-Abelian Gauge Dynamics in Two-Dimensional ${\cal N}=(2,2)$ Theories,''
  JHEP {\bf 0705}, 079 (2007)
  [hep-th/0609032].

\bibitem{Batyrev:1998mit}
  V.~V.~Batyrev, I.~Ciocan-Fontanine, B.~Kim and D.~van Straten,
  ``Mirror Symmetry and Toric Degenerations of Partial Flag
  Manifolds,''
  Acta Math. {\bf 184}, 1-39 (2000) [arXiv:9803108 [math.AG]].

\bibitem{Katz:1996fh}
  S.~H.~Katz, A.~Klemm and C.~Vafa,
  ``Geometric engineering of quantum field theories,''
  Nucl.\ Phys.\  B {\bf 497}, 173 (1997)
  [arXiv:hep-th/9609239].

\bibitem{Forbes:2005xt} 
  B.~Forbes and M.~Jinzenji,
  ``Extending the Picard-Fuchs system of local mirror symmetry,''
  J.\ Math.\ Phys.\  {\bf 46}, 082302 (2005)
  [hep-th/0503098].

\bibitem{Haghighat:2008gw} 
  B.~Haghighat, A.~Klemm and M.~Rauch,
  ``Integrability of the holomorphic anomaly equations,''
  JHEP {\bf 0810}, 097 (2008)
  [arXiv:0809.1674 [hep-th]].

\bibitem{Chiang:1999tz} 
  T.~M.~Chiang, A.~Klemm, S.~-T.~Yau and E.~Zaslow,
  ``Local mirror symmetry: Calculations and interpretations,''
  Adv.\ Theor.\ Math.\ Phys.\  {\bf 3}, 495 (1999)
  [hep-th/9903053].

\bibitem{Sharpe:2012ji} 
  E.~Sharpe,
  ``Predictions for Gromov-Witten invariants of noncommutative resolutions,''
  arXiv:1212.5322 [hep-th].

\bibitem{Griffiths:1978} 
  P.~Griffiths and J.~Harris,
  ``Principles of Algebraic Geometry,''
  Wiley,\ New York,\ 1978.

\bibitem{Borel:1958} 
  A.~Borel and F.~Hirzebruch,
  ``Characteristic classes and homogeneous spaces I,''
  Amer. J. Math. {\bf 80}, 458-538 (1958).

\bibitem{Haghighat:2008ut} 
  B.~Haghighat and A.~Klemm,
  ``Topological Strings on Grassmannian Calabi-Yau manifolds,''
  JHEP {\bf 0901}, 029 (2009)
  [arXiv:0802.2908 [hep-th]].




\end{thebibliography}
\end{document}